\documentclass[aps,pre,amsmath,amssymb,reprint,showkeys,showpacs,dvipdfmx]{revtex4-1}
\usepackage{amsmath}
\usepackage{amssymb}
\usepackage{bm}
\usepackage{dcolumn}
\usepackage[xdvi]{graphicx}
\usepackage{mathrsfs}
\makeatletter

\def\btt#1{\texttt{\@backslashchar#1}}%
\DeclareRobustCommand\bblash{\btt{\@backslashchar}}%
\makeatother

\newcommand{\bv}[1]{{\bm #1}}

\usepackage{version}	% required for `\comment' (yatex adctbded)
\begin{document}

\title{Divergence of Viscosity in Jammed Granular Materials: A Theoretical Approach}
\author{Koshiro Suzuki}
\address{Analysis Technology Development Center, Canon Inc., 30-2 Shimomaruko 3-chome, Ohta-ku, Tokyo 146-8501, Japan}
\author{Hisao Hayakawa\footnote{e-mail: hisao@yukawa.kyoto-u.ac.jp}}
\address{Yukawa Institute for Theoretical Physics, Kyoto University, Kitashirakawaoiwake-cho, Sakyo-ku, 
Kyoto 606-8502, Japan}

\begin{abstract}
A theory for jammed granular materials is developed with the aid of a
nonequilibrium steady-state distribution function.
The approximate nonequilibrium steady-state distribution function is
explicitly given in the weak dissipation regime by means of the
relaxation time.
The theory quantitatively agrees with the results of the molecular
dynamics simulation on the critical behavior of the viscosity
%in the vicinity of 
below the jamming point without introducing any fitting parameter.
\end{abstract}

\maketitle

%\section{Introduction}
{\it Introduction.--}
Description of granular rheology has been a long-term challenge for both
science and technology.
The problem extends to a vast range, from solid-like creep motion,
gas-like, to liquid-like phenomena~\cite{JNB1996}.
Similar to solid-liquid transitions, granular materials acquire rigidity
when the density exceeds a critical value~\cite{OLLN2002, OSLN2003,
OH2014, CSD2014}.
This phenomenon, referred to as the jamming transition, is universely
observed in disordered materials such as colloidal
suspensions~\cite{Pusey91}, emulsions, and foams~\cite{DW1994}, as well
as granular materials.
The jamming transition and its relation to the glass transition have
attracted much interest in the last two decades, and various aspects
have been revealed~\cite{LN1998, IBS2012, TS2010, PZ2010, LN2010}.
%
%[Discuss the features, attractiveness, and the universality of the
%jamming transition, citing review articles.]
%
%%%
%
In particular, characteristics in the vicinity of the jamming point,
including the critical scaling behavior, have been extensively
investigated by experiments and numerical simulations~\cite{OLLN2002,
OSLN2003, dCEPRC2005, WSNW2005, OT2007, H2007, H2008, OH2009-2,
OH2009-3, OH2011, TWRSH2010, BZCB2011, OT2012, OT2013, OH2014,
GDLW2015}.
%
%[Cite other papers of Olsson-Teitel etc.]
%
It has been shown that the shear stress, the pressure, and the granular
temperature can be expressed by scaling functions with exponents near
$\varphi \sim \varphi_{J}$, where $\varphi_{J}$ is the jamming
transition density.
The shear viscosity exhibits a form $\eta \sim
(\varphi_{J}-\varphi)^{-\lambda}$ with $\lambda \approx 2$ for
non-Browninan suspensions, foams, and emulsions~\cite{BGP2011, TAC2012,
ABH2012, LDW2012}, and a recent careful analysis demonstrated that
$\lambda$ is in the range between 1.67 and 2.55~\cite{KCIB2015}.
It seems that the exponent $\lambda$ for granular flows takes a larger
value than that for suspensions~\cite{OH2009-2, OH2009-3, OHL2010,
GDLW2015},
%
%[delete [CSS2012] if not appropriate.]
%
although a value in the range mentioned above has been reported as
well~\cite{H2008}.
However, these studies are based on numerical simulations or
phenomenologies without any foundation of a microscopic theory.

Even when we focus only on the flow properties below the jamming point
$\varphi_{\mathrm{J}}$, which can be tracked back to Bagnold's
work~\cite{Bagnold1954}, we have not yet obtained a complete set in
describing the rheological properties of dense granular flows.
One of the remarkable achievements is the extension of the
Boltzmann-Enskog kinetic theory to inelastic hard disks and
spheres~\cite{JR1985-1, JR1985-2, GD1999, L2005, SH2007, BP}.
However, it has been recognized that the kinetic theory breaks down at
densities with volume fraction $\varphi > \varphi_{f} =
0.49$~\cite{Louge2003, MN2005, MN2007, JB2010}, since there exists
correlated motions of grains.
%
%Empirical 
A modification of the kinetic theory has been proposed~\cite{CS2013},
but a microscopic theory is still absent.
Due to these situations, a microscopic liquid theory valid in the regime
$\varphi_{f} < \varphi < \varphi_{\mathrm{J}}$ has been desired.
One attempt to respond to this problem is the extension of the
mode-coupling theory (MCT)~\cite{Goetze} for dense granular liquids.
%
%This approach compensates the ignorance of the steady-state distribution
%function by solving the generalized Langevin equation for the transient
%density correlation function.
%%
%Although it reproduces the shear viscosity semi-quantitatively in the
%range $0.5 < \varphi < 0.60$, it requires a shift in $\varphi$ and
%predicts a non-exponential relaxation of the density correlation
%function, in contradiction to the result of numerical simulations.
%%
%Besides, it predicts a weaker divergence of the shear viscosity as
%$\varphi \to \varphi_{J}$ than the known results.
%
MCT has been applied to granular systems driven by Gaussian
noises~\cite{KSZ2010, SKZ2012, KSZ2013}.
It qualitatively predicts a liquid-glass transition, though the noise in
granular systems is non-Gaussian in general~\cite{ TWV2011, GPD2013,
KSSH2015-1, KSSH2015-2}.
MCT also has been applied to sheared dense granular
systems~\cite{HO2008, SH2013-3, SH2014}. 
There are three disadvantages of this approach:
(i) the shift of $\varphi$ is necessary to describe the divergence of
$\eta$. 
(ii) Because of the shift of $\varphi$, the jamming transition is not
correlated with the divergence of the first peak of the radial
distribution function.
(iii) It predicts a plateau in the density correlation function, which
is not observed in experiments nor in simulations~\cite{DMB2005, CC2009,
Kumaran2009}.

From these observations, it is crucial to obtain an explicit expression
of the steady-state distribution function to construct a theory for
dense granular liquids.
%
%, it is necessary to go beyond the Boltzmann-Enskog equation.
%
%The perturbative solution of the Boltzmann-Enskog equation \cite{CC70}
%is based on the expansion by spatial gradients of hydrodynamic fields.
%
For our purpose, we attempt to perform an expansion with respect to the
dissipation to obtain an approximate explicit expression of the
distribution function, valid in the weak dissipation regime for
frictionless granular flows.
Once the distribution is obtained explicitly, it is possible to
calculate the steady-state average for arbitrary observables.
%

%The organization of this letter is as follows: we first introduce the
%equation of motions for the particles, which is numerically solved in MD
%simulations.
%%
%Then, we derive the nonequilibrium steady-state distribution function
%valid in the weak dissipation regime.
%%
%The formulas for the steady-state average of the shear viscosity and the
%granular temperature are shown and compared to the results of MD
%simulations. 
%%
%We discuss the implications of this letter and conclude with some
%remarks. 
%%

%\section{Microscopic starting equations}
%\label{sec:microscopic-starting-point}
{\it Microscopic starting equations.--}
We consider a three-dimensional system of $N$ smooth granular particles
of mass $m$ and diameter $d$ in a volume $V$ subjected to stationary
shearing characterized by the shear-rate tensor $\bm{\dot\gamma}$.
We assume that each granular particle is a soft-sphere, and the contact
force acts only on the normal direction.
For a simple uniform shear with velocity along the $x$ axis and its
gradient along the $y$ axis, the shear-rate tensor is
$\dot\gamma_{\mu\nu} = \dot{\gamma} \delta_{\mu x} \delta_{\nu y}$
($\mu, \nu= x, y, z$) with a shear rate $\dot{\gamma}$.
It is postulated that the applied shear induces a homogeneous
streaming-velocity profile $\bm{\dot\gamma} \cdot \bm{r}$ at position
$\bm{r}$, assuming that no heterogeneity such as shear banding exists.
Thus, the equation of motion is given by
\begin{subequations}
\label{eq:Sllod}
\begin{eqnarray}
\bm{p}_{i}
&=&
m \left( \dot{\bm{r}}_{i} - \bm{\dot\gamma}\cdot\bm{r}_{i} \right),
\label{eq:Sllod-a}
\\
\dot{\bv{p}}_{i} 
&=& 
\bv{F}^{\rm (el)}_{i}+\bv{F}^{\rm (vis)}_i 
- 
\bm{\dot\gamma} \cdot \bv{p}_{i} ,
\label{eq:Sllod-b}
\end{eqnarray}
\end{subequations}
where $\bm{F}_i^{(\mathrm{el})}$ and $\bm{F}_i^{(\mathrm{vis})}$ are,
respectively, the elastic and the viscous contact forces acting on the
grain $i$.
Equation~(\ref{eq:Sllod}) is known as the Sllod equation, which is
equivalent to Newton's equation of motion under a uniform
shear~\cite{EM}.

The most essential feature of granular systems, in contrast to thermal
systems, is that the steady state is determined by the balance between
the viscous heating and the energy dissipation due to inelastic
collisions.
For sheared granular systems, this can be seen from the time derivative
of the Hamiltonian,
$
\mathcal{H}(\bm{\Gamma})
=
\sum_{i=1}^{N}
\bm{p}_{i}^2/(2m)
+
\sum_{i, j}{}^{'}
u(r_{ij}),
$
where $u(r_{ij})$ is the inter-particle potential depending on
$r_{ij}=|\bm{r}_i-\bm{r}_j|$, $\bm{\Gamma} = \{\bv{r}_{i},
\bv{p}_{i}\}_{i=1}^{N}$ is the phase-space coordinate, and
$\sum_{i,j}{}^{'}$ is the summation under the condition $i\ne j$.
Then $\dot{\mathcal{H}}(\bm{\Gamma})$ satisfies
\begin{eqnarray}
\dot{\mathcal{H}}(\bm{\Gamma}) 
=
-\dot\gamma V \sigma_{xy}(\bm{\Gamma})
-2\mathcal{R}(\bm{\Gamma}),
\label{eq:H0dot}
\end{eqnarray}
where
\begin{eqnarray}
\sigma_{\mu\nu}(\bm{\Gamma})
=
\frac{1}{V}
\sum_{i=1}^{N}
\left[
\frac{p_{i,\mu}p_{i,\nu}}{m}
+
r_{i,\nu}
\left(
F_{i,\mu}^{(\mathrm{el})} + F_{i,\mu}^{(\mathrm{vis})}
\right)
\right]
\label{eq:sigma}
\end{eqnarray}
is the stress tensor and 
\begin{eqnarray}
\mathcal{R}(\bm{\Gamma}) 
=
-\frac{1}{2}
\sum_{i=1}^{N}
\dot{\bm{r}}_{i}\cdot\bm{F}_{i}^{(\mathrm{vis})}
\label{eq:R}
\end{eqnarray}
corresponds to the Rayleigh's dissipation function~\cite{LL-Mech,
App}.
%
%The granular temperature is determined by the energy balance.
%
For granular systems with the interparticle dissipative force
proportional to the relative velocity, it is impossible to reduce the
dynamics as overdamped.
For later analysis, we assume that the contact forces
$\bm{F}_i^{(\mathrm{el})}=\sum_{j\ne i}\bm{F}_{ij}^{(\mathrm{el})}$ and
$\bm{F}_i^{(\mathrm{vis})}=\sum_{j\ne i}\bm{F}_{ij}^{(\mathrm{vis})}$
are, respectively, given by $\bm{F}_{ij}^{(\mathrm{el})}=\kappa\,
\Theta(d-r_{ij})(d-r_{ij})$ and $\bm{F}_{ij}^{(\mathrm{vis})}=-\zeta\,
\Theta(d-r_{ij}) (\bm{v}_{ij}\cdot\hat{\bm{r}}_{ij})\hat{\bm{r}}_{ij}$,
where $\Theta(x)=1$ for $x\ge 0$ and $\Theta(x)=0$ otherwise,
$\bm{r}_{ij}=\bm{r}_i-\bm{r}_j$, $\hat{\bm{r}}_{ij}=\bm{r}_{ij}/r_{ij}$,
and $\bm{v}_{ij}=\bm{v}_i-\bm{v}_j$ with the velocity of $i$-th particle
$\bm{v}_i$.
%
%The equation of motion, i.e. the Sllod equation Eq.~(\ref{eq:Sllod}),
%can be converted into the Liouville equation for
%$\bm{\Gamma}$~\cite{App}.
%

%\section{Steady-state distribution function}
%\label{subsec:SSD}
{\it Steady-state distribution function.--}
To address the distribution function for the nonequilibrium steady
state, we start from an equilibrium state at $t\to -\infty$ and evolve
the system with shear and dissipation.
Then, the system is expected to reach a steady state at $t=0$.
%
%%%%%
%
%From the Liouville equation, we expect that the nonequilibrium
%steady-state distribution function is a zero eigenfunction of the
%adjoint Liouvillian, $i\mathcal{L}^{\dagger}\,\rho(\bm{\Gamma},t=0)=0$.
%
Although it is impossible to derive an exact solution of the Liouville
equation, equivalent to Eq.~(\ref{eq:Sllod}), for the $6N$-dimensional
distribution function, it is possible to obtain an approximate solution
by perturbation, parallel to the method for the linearized Boltzmann
equation~\cite{R1970}.
In the perturbation for dense sheared granular systems, it is simple to
obtain the leading-order eigenfrequency of the relaxation towards the
steady state~\cite{App}.
Hence, we attempt to speculate an {\it approximate} steady-state
distribution, which we denote $\rho_{\mathrm{SS}}(\bm{\Gamma})$, by
applying an approximation which explictly utilizes the relaxation time.

For this purpose, we start from a formal but {\it exact} expression for
the distribution function \cite{EM},
\begin{equation}
\rho_{\mathrm{SS}}^{(\mathrm{ex})}(\bm{\Gamma})
=
\exp
\left[
\int_{-\infty}^{0} d\tau \,
\Omega_{\mathrm{eq}}(\bm{\Gamma}(-\tau))
\right]
\rho_{\mathrm{eq}}(\bm{\Gamma}(-\infty)),
\label{eq:rho_SS_ex}
\end{equation}
which is the steady-state solution of the Liouville equation.
Here,
$
\Omega_{\mathrm{eq}}(\bm{\Gamma}) 
=
\beta_{\mathrm{eq}}
\dot{\mathcal{H}}(\bm{\Gamma}) - \Lambda(\bm{\Gamma}) 
=
-\beta_{\mathrm{eq}}
\left[
\dot\gamma V \sigma_{xy}(\bm{\Gamma}) + 2\mathcal{R}(\bm{\Gamma})
\right] 
-\Lambda(\bm{\Gamma}) 
$
is the work function for $\rho_{\mathrm{eq}}(\bm{\Gamma}) =
e^{-\beta_{\mathrm{eq}}\mathcal{H}(\bm{\Gamma})}/\int d\bm{\Gamma} \,
e^{-\beta_{\mathrm{eq}}\mathcal{H}(\bm{\Gamma})}$ at temperature
$T_{\mathrm{eq}}=\beta_{\mathrm{eq}}^{-1}$, where
$\Lambda(\bm{\Gamma})=(\partial /\partial
\bm{\Gamma})\cdot\dot{\bm{\Gamma}}$ is the phase-space volume
contraction.
We approximate Eq.~(\ref{eq:rho_SS_ex}) by introducing the relaxation
time $\tau_{\mathrm{rel}}$ as
\begin{eqnarray}
\exp
\left[
\int_{-\infty}^{0} d\tau \,
\Omega_{\mathrm{eq}}(\bm{\Gamma}(-\tau))
\right]
\approx
e^{\tau_{\mathrm{rel}}\Omega_{\mathrm{SS}}(\bm{\Gamma})},
\label{eq:relaxation_time_approx}
\end{eqnarray}
which can be validated in the perturbation expansion of the Liouville
equation around the canonical distribution~\cite{App}.
In the perturbation, we non-dimensionalize all the quantities, where the
units of mass, length, and time are chosen as $m$, $d$, and
$\sqrt{m/\kappa}$, and introduce $\epsilon \equiv \zeta/\sqrt{\kappa
m}\ll 1$ as a perturbation parameter, which is related to the
restitution coefficient $e$ as $\epsilon \approx \sqrt{2}(1-e)/\pi$ for
$e\approx 1$.
We attach a star $*$ to the non-dimensionalized quantites, e.g. $t^* = t
\sqrt{\kappa/m}$.
%
%%%
%
Furthermore, we perform a scaling which leaves the steady-state
temperature~$T_{\mathrm{SS}}$, which is the ensemble average of
$\sum_{i=1}^{N} \bm{p}_{i}^2/(3Nm)$ at the steady state in the
dimensional unit, to be independent of $\epsilon$.
This indicates that the granular fluid keeps its motion in the limit
$\epsilon \to 0$.
From dimensional analysis, $T_{\mathrm{SS}}$ satisfies
$
T_{\mathrm{SS}} 
\sim
m^3 d^2 \dot\gamma^4/\zeta^2,
$
which leads to 
$
T^{*}_{\mathrm{SS}}
\sim
\epsilon^{-2}\, \dot\gamma^{*4}.
$
Hence, $\dot\gamma^{*}$ should satisfy
$
\dot\gamma^{*}
\sim
\epsilon^{1/2}.
$
We introduce a scaled shear rate $\tilde{\dot\gamma}$ as
$
\dot\gamma^{*}
=
\epsilon^{1/2}
\tilde{\dot\gamma},
$
where $\tilde{\dot\gamma}$ is independent of $\epsilon$.
We attach a tilde to the scaled quantities.
The relaxation time $\tau_{\mathrm{rel}}$ is evaluated from the
eigenfrequency of the perturbation expansion as
\begin{eqnarray}
\tau_{\mathrm{rel}}
=
\left[ \frac{2\sqrt{\pi}}{3}\epsilon\,
 \omega_{E}(T_{\mathrm{SS}})\right]^{-1}
\label{eq:tau_rel}
\end{eqnarray}
in the hard-core limit~\cite{App}, where
$\omega_{E}(T)=4\sqrt{\pi}\,n\sqrt{T/m}\, g_{0}(\varphi)d^2$ is the
Enskog frequency of collisions and $g_{0}(\varphi)$ is the first-peak
value of the radial distribution function.
%
%
% in Eq.~(\ref{eq:epsilon_def}).
%
In Eq.~(\ref{eq:relaxation_time_approx}), we have also introduced
\begin{equation}
\Omega_{\mathrm{SS}}(\bm{\Gamma}) 
=
-\beta_{\mathrm{SS}}
\hspace{-0.2em}
\left[
\dot\gamma V \sigma_{xy}^{(\mathrm{el})}(\bm{\Gamma})
+
2\Delta \mathcal{R}_{\mathrm{SS}}^{(1)}(\bm{\Gamma})
\right],
\label{eq:Omega_SS}
\end{equation}
where $\sigma_{xy}^{(\mathrm{el})}$ and $\Delta R_{\mathrm{SS}}^{(1)}$
are respectively given by~\cite{App}
\begin{eqnarray}
\sigma_{xy}^{(\mathrm{el})}(\bm{\Gamma}) 
&=&
\frac{1}{V}
\sum_{i=1}^{N}
\left[
\frac{p_{i,x} p_{i,y}}{m}
+
y_i F_{i,x}^{(\mathrm{el})}
\right],
\\
\Delta \mathcal{R}_{\mathrm{SS}}^{(1)}(\bm{\Gamma})
&=&
\mathcal{R}^{(1)}(\bm{\Gamma})
+
\frac{T_{\mathrm{SS}}}{2}
\Lambda(\bm{\Gamma}),
\\
\mathcal{R}^{(1)}(\bm{\Gamma})
&=&
\frac{\zeta}{4}
\sum_{i,j}{}^{'}
\hspace{-0.3em}
\left( \frac{\bm{p}_{ij}}{m}\cdot\hat{\bm{r}}_{ij}\right)^2
\Theta(d-r_{ij}).
\end{eqnarray}
Here, we ignore the contribution from the viscous shear stress, which is
a higher-order correction in the limit $\epsilon \to 0$.
To summarize, we obtain
\begin{eqnarray}
\rho_{\mathrm{SS}}(\bm{\Gamma})
=
\frac{e^{-I_{\mathrm{SS}}(\bm{\Gamma})}}
{\int d\bm{\Gamma} e^{-I_{\mathrm{SS}}(\bm{\Gamma})}},
\label{eq:rho_SS}
\end{eqnarray}
where
$
I_{\mathrm{SS}}(\bm{\Gamma})
=
\beta_{\mathrm{SS}}^{*} 
\mathcal{H}^{*}(\bm{\Gamma}) 
-\tilde{\tau}_{\mathrm{rel}} \tilde{\Omega}_{\mathrm{SS}}(\bm{\Gamma})
$
with
$
\tilde{\Omega}_{\mathrm{SS}}(\bm{\Gamma})
=
-\beta_{\mathrm{SS}}^{*}
\left[
\tilde{\dot\gamma} V^*
\tilde{\sigma}_{xy}^{(\mathrm{el})}(\bm{\Gamma})
+
2\Delta\tilde{\mathcal{R}}^{(1)}_{\mathrm{SS}}(\bm{\Gamma})
\right]
$.
We note that 
(i) the steady-state temperature
$T_{\mathrm{SS}}=\beta_{\mathrm{SS}}^{-1}$ appears in
Eq.~(\ref{eq:rho_SS}),
(ii) the steady-state average,
$
\langle \cdots \rangle_{\mathrm{SS}} 
\equiv
\int d\bm{\Gamma}\,
\rho_{\mathrm{SS}}(\bm{\Gamma})
\cdots,
$
is independent of the equilibrium temperature for
$\rho_{\mathrm{eq}}(\bm{\Gamma})$,
(iii) the problem reduces to an {"equilibrium"} 
one with an effective Hamiltonian
$\mathcal{H}_{\mathrm{eff}}(\bm{\Gamma}) =T_{\rm SS}I_{\rm
SS}(\bm{\Gamma})$ and the temperature $T_{\rm SS}$.
%
%%%
%
Because the nonequilibrium contribution is small, we further expand
$\rho_{\mathrm{SS}}(\bm{\Gamma})$ as
\begin{eqnarray}
\rho_{\mathrm{SS}}(\bm{\Gamma}) 
\approx
\frac{
e^{- \beta_{\mathrm{SS}}^{*} \mathcal{H}^{*}(\bm{\Gamma})}
\left[
1
+
\tilde{\tau}_{\mathrm{rel}}
\tilde{\Omega}_{\mathrm{SS}}(\bm{\Gamma})
\right]
}
{\mathcal{Z}}
\label{eq:rho_SS_expand}
\end{eqnarray}
with
$
\mathcal{Z}
\approx
\int d\bm{\Gamma} \,
e^{- \beta_{\mathrm{SS}}^{*} \mathcal{H}^{*}(\bm{\Gamma})}
\left[
1
+
\tilde{\tau}_{\mathrm{rel}}
\tilde{\Omega}_{\mathrm{SS}}(\bm{\Gamma})
\right].
$
%
%where we neglect terms of order $\mathcal{O}(\epsilon)$.
%
An approximate expression for $\langle
A(\bm{\Gamma})\rangle_{\mathrm{SS}}$ is obtained as
\begin{eqnarray}
\left\langle
A(\bm{\Gamma})
\right\rangle_{\mathrm{SS}}
&\approx&
\left\langle
A(\bm{\Gamma})
\right\rangle_{\mathrm{eq}}
+
\tilde{\tau}_{\mathrm{rel}}
\left\langle
A(\bm{\Gamma})
\tilde{\Omega}_{\mathrm{SS}}(\bm{\Gamma})
\right\rangle_{\mathrm{eq}},
\hspace{0.5em}
\label{eq:A_ave}
\end{eqnarray}
where
$
\left\langle
\cdots
\right\rangle_{\mathrm{eq}}
=
\int d\bm{\Gamma} \,
e^{- \beta_{\mathrm{SS}}^{*} \mathcal{H}^{*}(\bm{\Gamma})}
\cdots
$
is the average with respect to the canonical distribution at
$T_{\mathrm{SS}}$.
It should be noted that Eq.~(\ref{eq:A_ave}) is the result of an
exponential damping in the stress-stress correlation function in
the Green-Kubo formula.

So far $T_{\mathrm{SS}}$ is undetermined.
We attempt to determine $T_{\mathrm{SS}}$ by imposing the energy balance
\begin{equation}
\left\langle
\dot{\mathcal{H}}(\bm{\Gamma})
\right\rangle_{\mathrm{SS}}
=
-\dot{\gamma} V 
\left\langle
\sigma_{xy}(\bm{\Gamma})
\right\rangle_{\mathrm{SS}}
-2
\left\langle
\mathcal{R}(\bm{\Gamma})
\right\rangle_{\mathrm{SS}}
= 0.
\label{eq:ss}
\end{equation}
The explicit form of $T_{\mathrm{SS}}$ will be given in
Eq.~(\ref{eq:Tss}).
%
%Hence, the steady state is essentially characterized by the parameters,
%$(N, V, \dot\gamma)$.

%\subsection{Scaling for weak dissipation and shear}
%\label{subsec:scaling}
%
%%%
%

%\section{Shear viscosity and temperature}
%
{\it Shear viscosity and temperature.--}
Now we calculate the steady-state average of the shear stress and the
energy dissipation rate by Eq.~(\ref{eq:A_ave}) and derive an explicit
expression for $T_{\mathrm{SS}}$.
First, $\langle \sigma_{xy}(\bm{\Gamma})\rangle_{\mathrm{SS}}$ is
approximately given by
$
\left\langle
\tilde{\sigma}_{xy}(\bm{\Gamma})
\right\rangle_{\mathrm{SS}}
\approx
-
\tilde{\tau}_{\mathrm{rel}} \tilde{\dot\gamma}
\beta_{\mathrm{SS}}^{*} V^{*}
\left\langle
\tilde{\sigma}_{xy}^{(\mathrm{el})}(\bm{\Gamma})
\tilde{\sigma}_{xy}^{(\mathrm{el})}(\bm{\Gamma})
\right\rangle_{\mathrm{eq}}.
$
Similarly, the leading contribution gives
$
\left\langle
\tilde{\mathcal{R}}(\bm{\Gamma})
\right\rangle_{\mathrm{SS}}
\approx
\left\langle
\tilde{\mathcal{R}}^{(1)}(\bm{\Gamma})
\right\rangle_{\mathrm{eq}}
-
2\tilde{\tau}_{\mathrm{rel}} \beta_{\mathrm{SS}}^{*} 
\left\langle
\tilde{\mathcal{R}}^{(1)}(\bm{\Gamma})
\Delta\tilde{\mathcal{R}}_{\mathrm{SS}}^{(1)}(\bm{\Gamma})
\right\rangle_{\mathrm{eq}}.
$
%
%%%
%
\begin{figure*}[tbh]
%\begin{figure}[tbh]
%\begin{center}
\includegraphics[width=8.5cm]{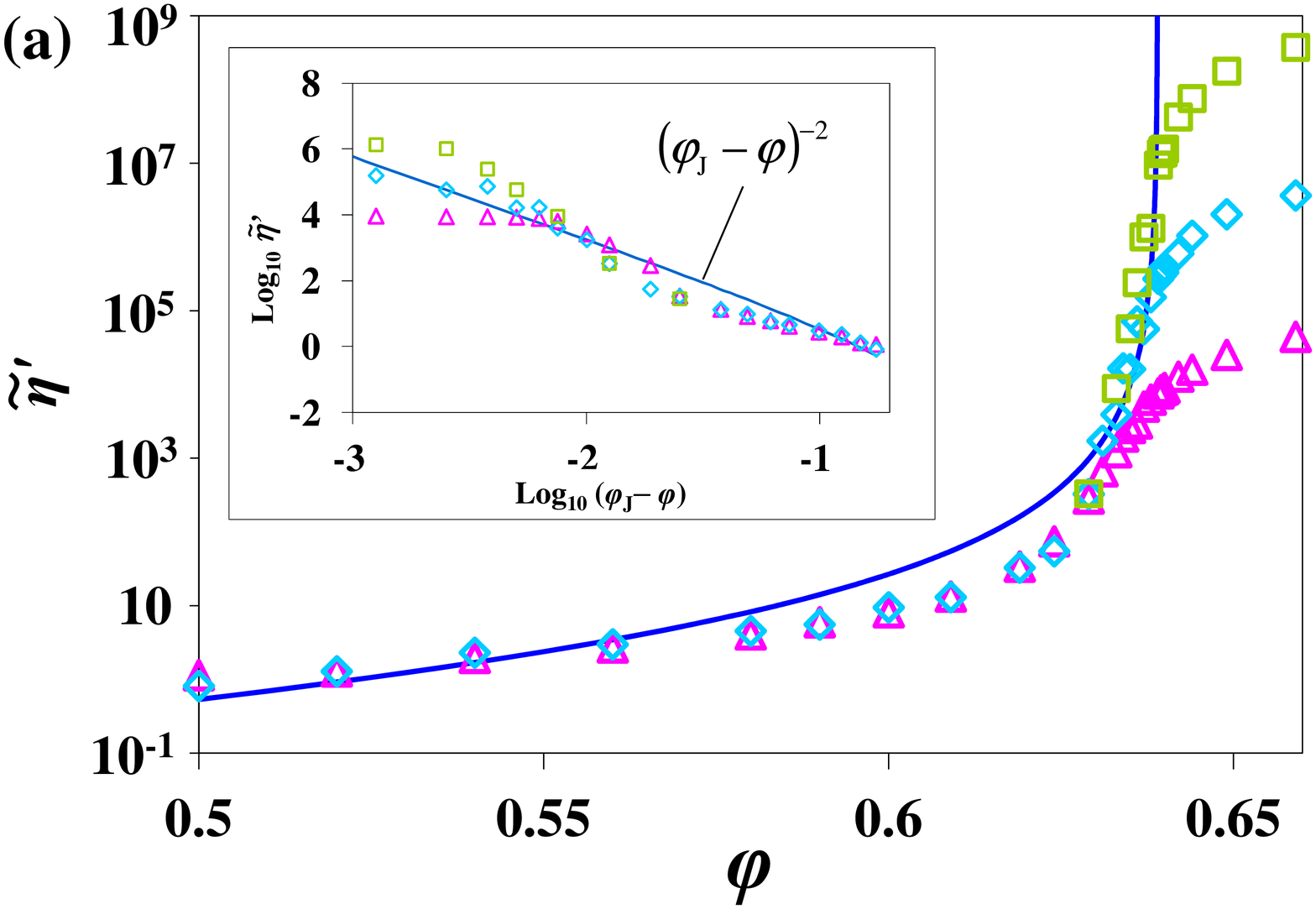} 
\includegraphics[width=8.5cm]{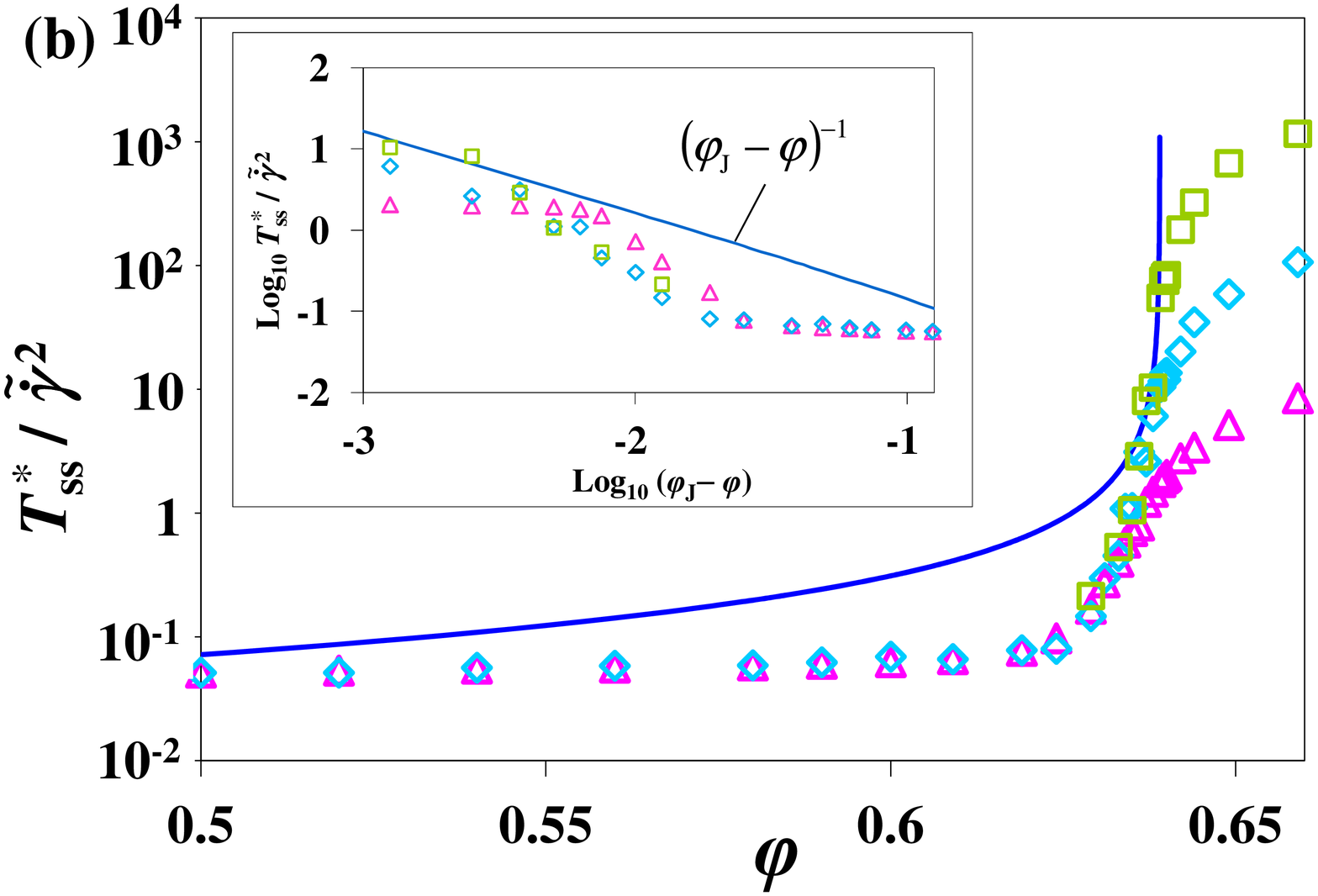} 
%\end{center} 
%\vspace{-2.0em}
\caption { (Color online) The density dependence of (a) the shear
viscosity $\tilde{\eta}'$ and (b) the granular temperature. 
The result of the theory is shown in (blue) solid line, while that for
the MD is shown in (red) triangles, (blue) diamonds, and (green)
rectangles for $\dot\gamma^{*}=10^{-3},10^{-4},10^{-5}$.
(Inset) The log-log plots for the results near $\varphi_{J}=0.639$.}
\label{Fig:VisTemp}
%\vspace{-1em}
\end{figure*}
%\end{figure}
%
%%%
%
Thus, we obtain the steady-state temperature from Eq.~(\ref{eq:ss}) as
\begin{eqnarray}
T_{\mathrm{SS}}^{*}
=
\frac{3\tilde{\dot\gamma}^2}{32\pi}
\frac{S}{R},
\label{eq:Tss}
\end{eqnarray}
where $S$ and $R$ are given by
$
S
=
1 + \mathscr{S}_{2}\, n^{*} g_{0}(\varphi) + \mathscr{S}_{3}\, n^{*2} g_{0}(\varphi)^2 +
\mathscr{S}_{4}\, n^{*3} g_{0}(\varphi)^3
$
and
$
R
=
n^{*}g_{0}(\varphi) \left[
\mathscr{R}_{2}' 
+ \mathscr{R}_{3}' \, n^{*}g_{0}(\varphi)
\right],
$
with $\mathscr{S}_{2}=2\pi/15$, $\mathscr{S}_{3}=-\pi^2/20$,
$\mathscr{S}_{4}=3\pi^3/160$, $\mathscr{R}_{2}' = -3/4$, and
$\mathscr{R}_{3}' = 7\pi/16$~\cite{App}.
We adopt the interpolation formula for hard spheres valid in the range
$\varphi_{f} < \varphi < \varphi_{J}$ ($\varphi_{f}=0.49$,
$\varphi_{J}=0.639$),
$g_{0}(\varphi)=g_{\mathrm{CS}}(\varphi_{f})(\varphi_{f}-\varphi_{J})/(\varphi-\varphi_{J})$,
where $g_{\mathrm{CS}}(\varphi)=(1-\varphi/2)/(1-\varphi)^3$ is the
formula by Carnahan and Starling valid at $\varphi < \varphi_{f}$
\cite{Torquato1995}.
Note that we adopt the Kirkwood approximation for many-body
correlations~\cite{App}.
We further obtain the expression for the shear stress,
\begin{equation}
\left\langle
\tilde{\sigma}_{xy}(\bm{\Gamma})
\right\rangle_{\mathrm{SS}}
=
%-
%\frac{3}{8\pi}
%\tilde{\dot\gamma}\,
%T_{\mathrm{SS}}^{* 1/2}
%\frac{S}{g_{0}(\varphi)}
%%\nonumber \\
%%%
%%&=&
%=
-
\frac{3\sqrt{6}}{64\pi^{3/2}}
\tilde{\dot\gamma}^2
\frac{S^{3/2}}{R^{1/2}g_{0}(\varphi)}.
\label{eq:sxy_ss}
\end{equation}
%
%%%
%
\begin{figure}[tbh]
%\begin{center}
\includegraphics[width=8.5cm]{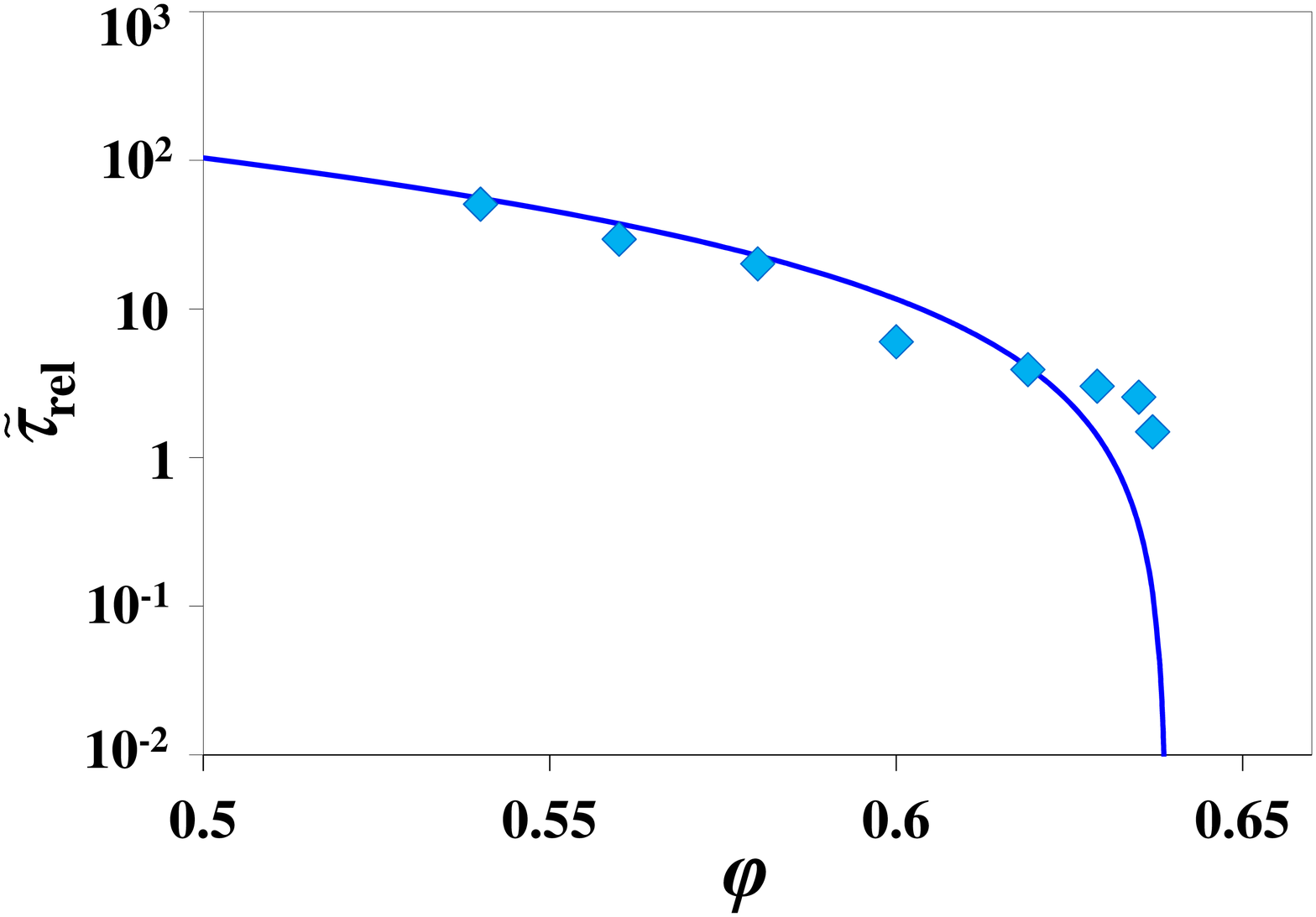} 
%\end{center} 
%\vspace{-2.0em}
\caption {(Color online) The density dependence of the relaxation time,
 $\tilde{\tau}_{\mathrm{rel}} = \epsilon \tau_{\mathrm{rel}}^{*}$. The
 result for the shear rate $\dot\gamma^{*}=10^{-4}$ is shown in (blue)
 solid line for the theory and (blue) diamonds for the MD simulation.  }
\label{Fig:tau_rel}
%\vspace{-1em}
\end{figure}
%
%%%
%
In the vicinity of the jamming point $\varphi_{J}$, $S$ and $R$ can be
approximated as $S\approx \mathscr{S}_{4}\, n^{*3}g_{0}(\varphi)^3$ and
$R\approx \mathscr{R}_{3}' \, n^{*2}g_{0}(\varphi)^2$, respectively.
This leads to the following expressions,
\begin{eqnarray}
T_{\mathrm{SS}}^{*} 
&\approx&
\frac{9\pi}{2240}\tilde{\dot\gamma}^2 n^{*} g_{0}(\varphi),
\\
\left\langle
\tilde{\sigma}_{xy}(\bm{\Gamma})
\right\rangle_{\mathrm{SS}}
&\approx&
- \frac{27\pi^{5/2}}{10240\sqrt{35}}
\tilde{\dot\gamma}^2
n^{* 7/2} g_{0}(\varphi)^{5/2},
\end{eqnarray}
from which we obtain 
\begin{eqnarray}
&&
T_{\mathrm{SS}}^{*}
\sim
(\varphi_{J}-\varphi)^{-1},
\label{eq:Tss_scaling}
\\
&&
\left\langle
\tilde{\sigma}_{xy}(\bm{\Gamma})
\right\rangle_{\mathrm{SS}}
\sim
(\varphi_{J}-\varphi)^{-5/2}.
\label{eq:sigma_xy_scaling}
\end{eqnarray}
From Eq.~(\ref{eq:sxy_ss}), % and (\ref{eq:sigma_xy_approx}), 
we obtain the shear viscosity $\eta^{*}=-\langle
\tilde{\sigma}_{xy}(\bm{\Gamma})\rangle_{\mathrm{SS}}/\tilde{\dot\gamma}
\sim (\varphi_{J}-\varphi)^{-5/2}$,
%
%\begin{eqnarray}
%\eta^{*}
%%&=&
%%\frac{3\sqrt{6}}{64\pi^{3/2}} 
%%\epsilon^{-1/2}\, \dot\gamma^{*} 
%%\frac{S^{3/2}}{R^{1/2}g_{0}(\varphi)}
%%\nonumber \\
%%%
%&\approx&
%\frac{27\pi^{5/2}}{10240\sqrt{35}}
%\epsilon^{-1/2} \dot\gamma^{*} n^{*7/2} g_{0}(\varphi)^{5/2}
%\sim
%(\varphi_{J}-\varphi)^{-5/2},
%\hspace{1.5em}
%\end{eqnarray}
%
or for $\tilde{\eta}' = -\langle
\tilde{\sigma}_{xy}(\bm{\Gamma})\rangle_{\mathrm{SS}}/(\tilde{\dot\gamma}
\sqrt{T_{\mathrm{SS}}^{*}}) \propto -\langle
\tilde{\sigma}_{xy}(\bm{\Gamma})\rangle_{\mathrm{SS}}/\tilde{\dot\gamma}^2$,
\begin{eqnarray}
\tilde{\eta}'
%&=&
%\frac{3}{8\pi} 
%\frac{S^{3/2}}{R^{1/2}g_{0}(\varphi)}
%%\nonumber \\
%%%
%%&\to&
%\approx
%\frac{9\pi^{2}}{1280}
%n^{*3} g_{0}(\varphi)^{2}
\sim
(\varphi_{J}-\varphi)^{-2}.
\hspace{1em}
\label{eq:eta_prime_scaling}
\end{eqnarray}
These results in
Eq.~(\ref{eq:Tss_scaling})--(\ref{eq:eta_prime_scaling}) are consistent
with the previous observations~\cite{BGP2011, TAC2012,
ABH2012, LDW2012, KCIB2015}.

%\section{Comparison with simulation}
%
{\it Comparison with simulation.--}
In order to verify the validity of the theory, we compare the theory
with the molecular dynamics (MD) simulation.
%
%We adopt the result for the nearly elastic monodisperse system,
%presented in Fig.~17 of Ref.~\cite{HO2009}.
%%
%The data presented in Ref.~\cite{HO2009} covers only the range $\varphi
%\geq 0.609$, so we have performed supplemental simulation for the range
%$0.50 \leq \varphi \leq 0.60$.
%
%%%%%
%In the MD simulation, the system is first thermalized in the absense of
%shear and dissipation at an initial temperature $T_{\mathrm{ini}}$.
%%
%Then, the shear and dissipation are switched on simultaneously, and the
%system is evolved until the temperature $T(t) = \sum_{i=1}^{N}
%\bm{p}_{i}(t)^2/(2m)$ starts to fluctuate around a steady value.
%%
%This value is identified as $T_{\mathrm{SS}}$.
%%
%Note that $T_{\mathrm{SS}}$ is determined solely by the balance of
%energy, i.e. $\dot\gamma$ and $\zeta$, and hence is independent of the
%choice of $T_{\mathrm{ini}}$.
%%%%%
%
The parameters in the MD are $N=2000$, $\epsilon = 0.018375$, and
$\dot\gamma^{*}=10^{-3}$, $10^{-4}$, $10^{-5}$.
This condition corresponds to $e=0.96$.

The shear viscosity $\tilde{\eta}'$ and $T_{\mathrm{SS}}$ are shown in
Fig.~\ref{Fig:VisTemp}, together with the results of the MD.
We also show the log-log plots near $\varphi_{J}$ as a function of
$\varphi_{J}-\varphi$ in the inset.
From the figure, we see that the theory agrees with the result of MD
simulation for $\varphi<\varphi_{J}$ quantitatively without introducing
any fitting parameter.
The agreement is refined as the shear rate is decreased, where the
hard-core limit is realized asymptotically.
The smeared divergence in the vicinity of the jamming point observed for
finite $\dot\gamma^{*}$ is a well-known feature of the soft-core MD.
We stress that the theory predicts the divergence of {\it both} the
shear viscosity and the granular temperature as $\varphi\to\varphi_{J}$,
in contrast to the kinetic theory of inelastic spheres, where the shear
viscosity behaves as
$
\tilde{\eta}'
\sim
(\varphi_{J}-\varphi)^{-1}$
and $T_{\mathrm{SS}}$ remains finite~\cite{GD1999}.
On the other hand, in the MD, the shear viscosity behaves as
$
\tilde{\eta}'
\sim
(\varphi_{J}-\varphi)^{-2}$,
in accordance with the theory.
%
%Furthermore, the theory also predicts the magnitudes relatively well.
%%, though there is some discrepancy in the temperature.
%
We note that the agreement between the theory and MD is relatively poor
for $T_{\mathrm{SS}}$, though we have not clarified its reason.

The relaxation time, Eq.~(\ref{eq:tau_rel}), is shown in
Fig.~\ref{Fig:tau_rel}, together with the result of the MD.
The result of the MD is extracted from fitting by an exponential
function for the transient data of the temperature relaxing to the
steady-state.
We see that Eq.~(\ref{eq:tau_rel}) is quantitatively valid for $\varphi
< 0.63$.

%\section{Discussions}
%
{\it Discussions.--}
From Eqs.~(\ref{eq:Tss}) and (\ref{eq:sxy_ss}), we see that the theory
is subjected to the Bagnold scaling.
The result of MD shows that the discrepancy from the Bagnold scaling
becomes significant for $\varphi > 0.635$.
%
%This is also true for the relaxation time.
%
Hence, there is room for improving the theory to cover the non-Bagnold
regime.
%
%%%
%
From the phenomenological scaling of jammed granules, the viscosity
exhibits $\eta \sim |\varphi_{J}-\varphi|^{y_{\phi}(1-2/y_{\gamma})}$,
where $y_{\phi}$ and $y_{\gamma}$ are the scaling exponents for
$\sigma_{xy}\sim (\varphi-\varphi_J)^{y_\phi}$ for $\varphi>\varphi_J$
and $\sigma_{xy}\sim \dot\gamma^{y_\gamma}$ at $\varphi =
\varphi_J$~\cite{OH2009-2}.
If we assume $y_{\phi}=1$ as in Refs.~\cite{OLLN2002, OSLN2003,
OH2009-2, OH2009-3}, we have $y_{\gamma}=4/7$, which is close to the
value of Ref.~\cite{H2008}
\footnote{ Although the scaling of $\sigma_{xy}$ must be
$(\varphi-\varphi_J)^{1/2}$ if the contact network is unchanged for
small strain $\gamma$, the contact network should be rearranged under a
plane shear, because $\gamma=\dot\gamma t$ is not small even for small
$\dot\gamma$ in the steady state in the limit $t\to \infty$. If $\gamma$
is not small, $\sigma_{xy}$ satisfies $y_\phi=1$~\cite{OH2014}}
.
For strongly dissipative situations, higher-order terms might alter the
exponents of the divergences.
Such a contribution will be discussed elsewhere. 

%\clearpage
%\section{Summary and Concluding Remarks}
%\label{sec:conclusion}
{\it Concluding Remarks.--}
We have developed a theory for jammed frictionless granular particles
subjected to a uniform shear with the aid of an approximate
nonequilibrium steady-state distribution function, and have shown that
it remarkably agrees with the result of the MD simulation below the
jamming point without introducing any fitting parameter.
There are many future tasks for the application of our theory, such as
the emergence of the shear modulus above the jamming
point~\cite{OLLN2002, OSLN2003, OH2014, CSD2014}, the
effect of friction for grains where the discontinuous shear thickening
appears~\cite{SMMD2013}, the drag force acting on the pulling
tracer~\cite{KD2007, TFO2010, KD2013, TO2014, TH2015}, etc.
Moreover, we should stress that the framework of our theory is quite
generic.
Indeed, we believe that the divergence of the viscosity for colloidal
suspensions, $\eta \sim (\varphi_{J}-\varphi)^{-2}$~\cite{Brady1993}, can
be understood by our framework.
Therefore, the theory is expected to be applicable to a wide variety of
phenomena in nonequilibrium processes.
%

%\begin{acknowledgments}
%
%\begin{comment}
%{\it Acknowledgements.--}
%
The authors are grateful to S.-H. Chong and M. Otsuki for their
contributions in the early stage of this work and extensive discussions.
They also thank M. Fuchs, M. E. Cates, M. Wyart, K. Saitoh, A. Ikeda,
and K. Kanazawa for fruitful discussions and comments, and M. Otsuki and
S. Takada for providing the prototype of the program for the MD
simulation.
% , and S. Torquato for informing us Ref.~\cite{TDS2003}.
%
This work was partially supported by Grant-in-Aids for scientific
research (Grant Nos. 25287098).
The MD simulation for this work has been carried out at the computer
facilities at the Yukawa Institute for Theoretical Physics, Kyoto
University.

%\end{comment}
%
%\end{acknowledgments}

\appendix

\section{Microscopic starting equations}
\label{app:sec:microscopic}

We introduce the system we consider, i.e. a three-dimensional system of
$N$ smooth granular particles of mass $m$ in a volume $V$ subjected to
stationary shearing characterized by the shear-rate tensor
$\dot\gamma_{\mu\nu}=\dot\gamma \delta_{\mu x}\delta_{\nu y}$.
The time evolution of the system is determined by the Newton's equation
of motion,
\begin{eqnarray}
m \ddot{\bv{r}}_{i} = \bv{F}_{i}^{\rm (el)} + \bv{F}_{i}^{\rm (vis)}
\hspace{0.5em}
(i=1,\cdots,N),
\label{eq:Newton}
\end{eqnarray}
under a suitable boundary condition such as the Lees-Edwards boundary
condition~\cite{EM}, accounting for the stationary shearing.
Here, $\bv{r}_{i}$ refers to the position of the $i$-th particle, and
the dot denotes the time derivative.
$\bv{F}^{\rm (el)}_{i} = \sum_{j}' \bv{F}_{ij}^{\rm (el)}$ is the
conservative force, and is given by a sum $\sum_j'$ of forces exerted on
the $i$-th particle by other particles,
\begin{eqnarray}
\bv{F}_{ij}^{\rm (el)}
=
-\frac{\partial u(r_{ij})}{\partial \bv{r}_{ij}}
=
\Theta(d-r_{ij})f(d-r_{ij})\hat{\bv{r}}_{ij}.
\label{eq:elastic-force}
\end{eqnarray}
Here, $d$ is the diameter of the particle; $u(r)$ is the pair potential;
$\bv{r}_{ij}=\bv{r}_{i} - \bv{r}_{j}$, $r_{ij}=|\bv{r}_{ij}|$,
$\hat{\bv{r}}_{ij}=\bv{r}_{ij}/r_{ij}$; and $\Theta(x)$ is the
Heaviside's step function, which is 1 for $x>0$ and 0 otherwise.
Although a realistic elastic force might be Hertzian, where $f(x)\propto
x^{3/2}$ for a three-dimensional system, we adopt the linear spring
model, $f(x) = \kappa x$ ($\kappa > 0$), for simplicity.
$\bv{F}^{\rm (vis)}_{i} = \sum_{j}' \bv{F}^{\rm (vis)}_{ij}$ 
denotes the viscous dissipative force which is represented by a sum
of pairwise contact forces, 
\begin{eqnarray}
\bv{F}^{\rm (vis)}_{ij} 
=
-
\zeta
\Theta(d-r_{ij})
\hat{\bm{r}}_{ij}
(\bm{v}_{ij}\cdot\hat{\bv{r}}_{ij}).
\label{eq:viscous-force}
\end{eqnarray}
Here, $\bm{v}_{ij} \equiv \bv{v}_{i} - \bv{v}_{j}$ with $\bv{v}_{i} =
\dot{\bv{r}}_{i}$ refers to the relative velocity of colliding
particles, and $\zeta$ ($>0$) is a viscous constant corresponding to the
harmonic potential.
This model for sheared granular systems has been widely adopted in
computer-simulation studies.
Notice that the equation of motion is not invariant under the
time-reversal map, since $\bv{F}^{\rm (vis)}_{ij}$ changes sign for $\{
\bv{r}_{i}, \bv{v}_{i} \} \to \{ \bv{r}_{i}, -\bv{v}_{i} \}$.

Instead of Eq.~(\ref{eq:Newton}), we consider the following equation of
motion valid for dense sheared granular systems,
\begin{subequations}
\label{app:eq:Sllod}
\begin{eqnarray}
\bm{p}_{i}
&=&
m \left( \dot{\bm{r}}_{i} - \bm{\dot\gamma}\cdot\bm{r}_{i}\right),
\label{app:eq:Sllod-a}
\\
\dot{\bv{p}}_{i} 
&=& 
\bv{F}^{\rm (el)}_{i}+\bv{F}^{\rm (vis)}_i 
- 
\bm{\dot\gamma} \cdot \bv{p}_{i} ,
\label{app:eq:Sllod-b}
\end{eqnarray}
\end{subequations}
which is called the Sllod equation~\cite{EM}.
Here, $\bv{p}_{i}$ is the fluctuation of momentum around the
steady-shear momentum and referred to as the peculiar, or thermal,
momentum.
The use of the Sllod equation is restricted to uniform shear, which is
in general not realized in realistic granular systems.
However, it is a valid idealization in the high-density regime or for
the flow on an inclined plane~\cite{SEG2001, SLG2003, MN2005}, where the
profile of the shear velocity is well approximated as linear except for
the region near the boundary.
Hence, we adopt the Sllod equation to address the rheology of dense
sheared granular liquids.

The internal energy of the system is given by
\begin{eqnarray}
\mathcal{H}(\bm{\Gamma})
=
\sum_{i=1}^{N}
\frac{\bm{p}_{i}^2}{2m}
+
\sum_{i, j}{}^{'}
u(r_{ij}),
\end{eqnarray}
where $\bm{\Gamma} = \{\bv{r}_{i}, \bv{p}_{i}\}_{i=1}^{N}$ is the
phase-space coordinate and $\sum_{i,j}{}^{'}$ is the summation under the
condition $i\ne j$.
The rate of change of the internal energy
$\dot{\mathcal{H}}(\bm{\Gamma})$ can be calculated, together with the
Sllod equation, Eq.~(\ref{app:eq:Sllod}), as
\begin{eqnarray}
\dot{\mathcal{H}}
&=&
\sum_{i=1}^{N}
\frac{\bm{p}_{i}}{m}
\cdot
\dot{\bm{p}}_{i}
+
\sum_{i,j}{}^{'}
\frac{\partial u(r_{ij})}{\partial \bm{r}_{i}}
\cdot
\dot{\bm{r}}_{i}
\nonumber \\
&=&
\sum_{i=1}^{N}
\frac{\bm{p}_{i}}{m}
\hspace{-0.2em}
\cdot
\hspace{-0.2em}
\left[
\bm{F}_{i}^{(\mathrm{el})}
\hspace{-0.2em}
+
\hspace{-0.2em}
\bm{F}_{i}^{(\mathrm{vis})}
\hspace{-0.3em}
-
\hspace{-0.2em}
\bm{\dot\gamma}\cdot\bm{p}_{i}
\right]
%\nonumber \\
%%
%&&
\hspace{-0.2em}
-
\hspace{-0.2em}
\sum_{i=1}^{N}
\bm{F}_{i}^{(\mathrm{el})}
\hspace{-0.3em}
\cdot
\hspace{-0.2em}
\left[
\frac{\bm{p}_{i}}{m}
+
\bm{\dot\gamma}\cdot\bm{r}_{i}
\right]
\nonumber \\
&=&
-\dot\gamma
\sum_{i=1}^{N}
\left[
\frac{p_{i,x} p_{i,y}}{m}
+
y_{i} F_{i,x}^{(\mathrm{el})}
\right]
+
\sum_{i=1}^{N}
\frac{\bm{p}_{i}}{m}\cdot\bm{F}_{i}^{(\mathrm{vis})}
\nonumber \\
&=&
-\dot\gamma
\sum_{i=1}^{N}
\left[
\frac{p_{i,x} p_{i,y}}{m}
+
y_{i} 
\left(
F_{i,x}^{(\mathrm{el})}
+
F_{i,x}^{(\mathrm{vis})}
\right)
\right]
+
\sum_{i=1}^{N}
\dot{\bm{r}}_{i}\cdot\bm{F}_{i}^{(\mathrm{vis})}
\nonumber \\
&=&
-\dot\gamma V
\sigma_{xy}
-2\mathcal{R}.
\label{eq:Hdot}
\end{eqnarray}
Here, $\sigma_{xy}(\bm{\Gamma})$ is the $xy$-component of the stress
tensor,
\begin{eqnarray}
\sigma_{\mu\nu}(\bm{\Gamma})
=
\frac{1}{V}
\sum_{i=1}^{N}
\left[
\frac{p_{i,\mu}p_{i,\nu}}{m}
+
r_{i,\nu}
\left(
F_{i,\mu}^{(\mathrm{el})} + F_{i,\mu}^{(\mathrm{vis})}
\right)
\right],
\label{app:eq:sigma}
\end{eqnarray}
and
\begin{eqnarray}
\mathcal{R}(\bm{\Gamma}) 
&=&
-\frac{1}{2}
\sum_{i=1}^{N}
\dot{\bm{r}}_{i}\cdot\bm{F}_{i}^{(\mathrm{vis})}
=
-\frac{1}{4}
\sum_{i, j}{}^{'}
\bm{v}_{ij}\cdot\bm{F}_{ij}^{(\mathrm{vis})}
\nonumber \\
&=&
\frac{\zeta}{4}
\sum_{i,j}{}'\Theta(d-r_{ij})(\bm{v}_{ij}\cdot\hat{\bm{r}}_{ij})^2
\label{app:eq:R}
\end{eqnarray}
corresponds to the Rayleigh's dissipation function~\cite{LL-Mech}.
Note that we have utilized $\bm{F}_{i}^{(\mathrm{vis})}=\sum_{j\neq
i}\bm{F}_{ij}^{(\mathrm{vis})}$ and
$\bm{F}_{ij}^{(\mathrm{vis})}=-\bm{F}_{ji}^{(\mathrm{vis})}$ in the last
equality.
The first term on the right hand side of Eq.~(\ref{eq:Hdot}) is the
work rate of shear exerted on the system and the second term is the
energy dissipation rate due to inelastic collisions.
In the steady state, the balance between these two are realized, i.e.
\begin{eqnarray}
\dot{\mathcal{H}}(\bm{\Gamma})
=
-\dot\gamma V \sigma_{xy}(\bm{\Gamma})
-2\mathcal{R}(\bm{\Gamma})
=0.
\label{app:eq:ss}
\end{eqnarray}

\section{Liouville equation}
\label{app:Liouville}

The equation of motion, i.e. the Sllod equation
Eq.~(\ref{app:eq:Sllod}), can be converted into the Liouville
equation~\cite{EM}.
For hard-sphere granular materials, it is conventional to adopt the
pseudo-Liouville equation \cite{BDS1997, DBB2008, BDB2008}.
However, we start with the Liouville equation for soft spheres and take
the hard-core limit later, since there is a subtlety in expressing the
shear stress and the dissipation function via the pseudo-Liouvillian.
We present the Liouville equation for uniformly sheared granular
systems in the following. 
It closely follows the formulation of Ref.~\cite{EM}, where similar
explicit expressions can also be found in Refs.~\cite{COH2010-3,
HO2013}. 

The time evolution of phase variables, which do not explicitly depend on
time and whose time dependence comes solely from that of the phase-space
point $\bm{\Gamma} = \{\bv{r}_{i}, \bv{p}_{i}\}_{i=1}^{N}$, is
determined by
\begin{equation}
\frac{d}{dt} A(\bv{\Gamma}(t)) 
=
\dot{\bv{\Gamma}} \cdot 
\frac{\partial}{\partial \bv{\Gamma}}
A(\bv{\Gamma}(t)) \equiv
i {\cal L} A(\bv{\Gamma}(t)).
\label{eq:Liouville_obs}
\end{equation}
The operator $i{\cal L}$ is referred to as the Liouvillian.  For
Eq.~(\ref{app:eq:Sllod}), we have
\begin{equation}
i {\cal L} 
= 
i {\cal L}^{\rm (el)} + i {\cal L}_{\dot{\gamma}} + i {\cal L}^{\rm (vis)}, 
\label{eq:iL-Sllod}
\end{equation}
where the elastic part ($i {\cal L}^{\rm (el)}$), the shear part $(i
{\cal L}_{\dot{\gamma}})$, and the viscous part $(i {\cal L}^{\rm
(vis)})$ are, respectively, given by
\begin{eqnarray}
i {\cal L}^{\rm (el)} &=& \sum_{i}
\Bigl[ \,
  \frac{\bv{p}_{i}}{m} \cdot \frac{\partial}{\partial \bv{r}_{i}} +
  \bv{F}_{i}^{\rm (el)} \cdot \frac{\partial}{\partial \bv{p}_{i}} \,
\Bigr],
\label{eq:iLel}
\\
i {\cal L}_{\dot{\gamma}} &=& \sum_{i}
\Bigl[ \,
  (\bm{\dot\gamma} \cdot \bv{r}_{i}) \cdot  
  \frac{\partial}{\partial \bv{r}_{i}} -
  (\bm{\dot\gamma} \cdot \bv{p}_{i}) \cdot  
  \frac{\partial}{\partial \bv{p}_{i}} \,
\Bigr],
\label{eq:iL-dot-gamma}
\\
i {\cal L}^{\rm (vis)} &=& \sum_{i}
  \bv{F}_i^{\rm (vis)} \cdot 
\frac{\partial}{\partial \bv{p}_{i}}.
\label{eq:iLvis}
\end{eqnarray}
The formal solution to Eq.~(\ref{eq:Liouville_obs}) can be written in
terms of the propagator $\exp( i {\cal L} t)$ as
\begin{equation}
A(\bv{\Gamma}(t)) =
\exp(i {\cal L} t)
A(\bv{\Gamma}).
\label{eq:p-propagator}
\end{equation}
Hereafter, the absence of the argument $t$ implies that associated
quantities are evaluated at $t=0$, e.g. $\bm{\Gamma}=\bm{\Gamma}(0)$.

The Liouville equation for the nonequilibrium phase-space distribution
function $\rho(\bv{\Gamma},t)$ is given by
\begin{eqnarray}
\frac{\partial \rho(\bv{\Gamma},t)}{\partial t} 
&=&
- \frac{\partial}{\partial \bv{\Gamma}} \cdot
\left[ \, \dot{\bv{\Gamma}} \rho(\bv{\Gamma},t) \, \right]
%\nonumber \\
%&=&
=
- \left[ \, \dot{\bv{\Gamma}} \cdot 
       \frac{\partial}{\partial \bv{\Gamma}} +
       \Lambda(\bv{\Gamma}) \,
\right] \rho(\bv{\Gamma},t) 
\nonumber \\
&\equiv&
- i {\cal L}^{\dagger} \rho(\bv{\Gamma},t),
\label{eq:Liouville_df}
\end{eqnarray}
where
$i\mathcal{L}^{\dagger}(\bm{\Gamma})=i\mathcal{L}(\bm{\Gamma})+\Lambda(\bm{\Gamma})$
is referred to as the adjoint Liouvillian.
Here, $\Lambda(\bv{\Gamma})$ denotes the phase-space contraction
factor~\cite{EM} which is defined by
\begin{equation}
\Lambda(\bv{\Gamma}) \equiv
\frac{\partial}{\partial \bv{\Gamma}} \cdot
\dot{\bv{\Gamma}}
=\sum_{i}
\Bigl( \,
  \frac{\partial}{\partial \bv{r}_{i}} \cdot \dot{\bv{r}}_{i} +
  \frac{\partial}{\partial \bv{p}_{i}} \cdot \dot{\bv{p}}_{i} \,
\Bigr)  .
\label{eq:Lambda}
\end{equation}
For the Sllod equation Eq.~(\ref{app:eq:Sllod}), one finds
\begin{equation}
\Lambda(\bv{\Gamma}) 
=
-\frac{\zeta}{m}\sum_{i,j}{}'\, \Theta(d - r_{ij}) < 0.
\label{eq:Sllod-Lambda}
\end{equation}
The formal solution of the Liouville equation (\ref{eq:Liouville_df})
reads
\begin{equation}
\rho(\bv{\Gamma},t) = \exp( - i {\cal L}^{\dagger} t) \,
\rho(\bv{\Gamma},0).
\label{eq:f-propagator}
\end{equation}
%
%
%%%%%%%%%%%%%%%%%%
%%%%%%%%%%%%%%%%%%
\begin{comment}
%
From Eqs.~(\ref{eq:Liouville_obs}) and (\ref{eq:Liouville_df}) we
readily obtain the relation
%
\begin{equation}
i{\cal L}^{\dagger}(\bv{\Gamma}) = 
i{\cal L}(\bv{\Gamma}) + \Lambda(\bv{\Gamma}).
\label{eq:relation-Liouville-operators}
\end{equation}
%
%%%
%
The operators $i{\cal L}$ and $i{\cal L}^{\dagger}$ satisfy the adjoint
relation
%
\begin{equation}
\int d\bv{\Gamma} \,
[ i {\cal L} A(\bv{\Gamma}) ] \, B(\bv{\Gamma}) =
- \int d\bv{\Gamma} \,
A(\bv{\Gamma}) \,
[ i {\cal L}^{\dagger} B(\bv{\Gamma})].
\label{eq:adjoint-property}
\end{equation}
%
This property can be proved from the integration by parts. 
%
By a repeated use of this property, the following relation for the
propagators can be derived~\cite{EM}:
%
\begin{equation}
\int d\bv{\Gamma} \,
[ e^{i {\cal L} t} A(\bv{\Gamma})] \, B(\bv{\Gamma}) =
\int d\bv{\Gamma} \,
A(\bv{\Gamma}) \, [ e^{- i {\cal L}^{\dagger} t} B(\bv{\Gamma})].
\label{eq:unrolling}
\end{equation} 
%
\end{comment}
%%%%%%%%%%%%%%%%%%%%%%
%%%%%%%%%%%%%%%%%%%%%%

\section{The perturbation expansion of the Liouville equation}
\label{app:tauH}

We attempt to derive the eigenfrequencies of the distribution function
for dense sheared granular liquids by means of a perturbation expansion
of Eq.~(\ref{eq:Liouville_df}).
In particular, we attempt to construct a Rayleigh-Schr{\" o}dinger
perturbation theory, where the dissipation and the shear are treated as
perturbations.

In Eq.~(\ref{eq:Liouville_df}), we consider the Laplace transform of
$\rho(\bm{\Gamma},t)$, i.e.
\begin{eqnarray}
\Psi_{n}(\bm{\Gamma})
=
\int_{-\infty}^{0} dt \,
e^{-z_{n} t}
\rho(\bm{\Gamma},t),
\end{eqnarray}
which satisfies
\begin{eqnarray}
i\mathcal{L}^{\dagger}(\bm{\Gamma})
\Psi_{n}(\bm{\Gamma})
=
-z_{n}
\Psi_{n}(\bm{\Gamma}).
\label{eq:LiouvilleEq}
\end{eqnarray}
Here, $n$ is an index for the Laplace modes, which is continuous or
discrete.
This is an eigenvalue equation for the adjoint Liouvillian.
The distribution function is given by
\begin{eqnarray}
\rho(\bm{\Gamma},t)
=
\sum_{n=0}^{\infty}
e^{z_{n}t} 
\Psi_{n}(\bm{\Gamma}),
\end{eqnarray}
where the summation over $n$ is an integral for continuous modes.

In order to perform a perturbation expansion, we first
non-dimensionalize the observables by choosing the unit of mass, length,
and time as $m$, $d$, and $\sqrt{m/\kappa}$, and introduce an
infinitesimal parameter
\begin{eqnarray}
\epsilon 
=
\frac{\zeta}{\sqrt{\kappa m}} \ll 1.
\end{eqnarray}
Note that the restitution coefficient $e$ is related to $\epsilon$ as
$e=\exp\left[ -\zeta t_{c}/m \right]$ via $\zeta t_{c}/m = \pi\,
\epsilon/\sqrt{2(1-\epsilon^2)}$, where
$t_{c}=\pi/\sqrt{2\kappa/m-(\zeta/m)^2}$ is the duration of contact of
the spheres \cite{OHL2010}.
For $\epsilon \ll 1$, the normalized energy dissipation rate $1-e^2$ can
be approximated as $1-e^2 \approx 2\zeta t_{c}/m = \sqrt{2}\pi \epsilon
+ \mathcal{O}(\epsilon^3) \approx \sqrt{2}\pi \epsilon$.
Together with $1-e^2 \approx 2(1-e)$, we have $\epsilon \approx
\sqrt{2}(1-e)/\pi$ for $e\approx 1$.
We attach a star $*$ to the non-dimensionalized quantites, e.g. $t^* = t
\sqrt{\kappa/m}$.
Furthermore, we perform a scaling which leaves the steady-state granular
temperature $T_{\mathrm{SS}}$, which is defined by $\int d\bm{\Gamma}\,
\rho_{\mathrm{SS}}(\bm{\Gamma}) \sum_{i=1}^{N} \bm{p}_{i}^2/(3Nm)$,
where $\rho_{\mathrm{SS}}(\bm{\Gamma})$ is the steady-state value of
$\rho(\bm{\Gamma},t)$, to be independent of $\epsilon$.
This is because we are interested in the situation where the granular
fluid keeps a meaningful fluid motion in the limit $\epsilon \to 0$.
From dimensional analysis, $\langle
\sigma_{xy}(\bm{\Gamma})\rangle_{\mathrm{SS}}$ and $\langle
\mathcal{R}(\bm{\Gamma}) \rangle_{\mathrm{SS}}$ scale as
$
\left\langle
\sigma_{xy}(\bm{\Gamma})
\right\rangle_{\mathrm{SS}} 
\sim
\dot\gamma
\sqrt{mT_{\mathrm{SS}}}/d^2
$
and
$
\left\langle
\mathcal{R}(\bm{\Gamma})
\right\rangle_{\mathrm{SS}}
\sim
\zeta
T_{\mathrm{SS}}/m
$.
From the energy balance, Eq.~(\ref{app:eq:ss}), $T_{\mathrm{SS}}$ scales
as
$
T_{\mathrm{SS}} 
\sim
m^3 d^2 \dot\gamma^4/\zeta^2,
$
which leads to 
$
T^{*}_{\mathrm{SS}}
\sim
\epsilon^{-2}\, \dot\gamma^{*4}.
$
Thus, if we require $T^{*}_{\mathrm{SS}}$ to be independent of
$\epsilon$, $\dot\gamma^{*}$ should scale as
\begin{eqnarray}
\dot\gamma^{*}
\sim
\epsilon^{1/2}.
\end{eqnarray}
To be specific, we introduce a scaled shear rate $\tilde{\dot\gamma}$ as
\begin{eqnarray}
\dot\gamma^{*}
=
\epsilon^{1/2}
\tilde{\dot\gamma},
\end{eqnarray}
where $\tilde{\dot\gamma}$ is independent of $\epsilon$.
We attach a tilde to the scaled quantities.
%
%%%
%
Although implicit, it should be kept in mind that the anisotropic shear
stress $\sigma_{xy}(\bm{\Gamma})$ is proportional to the shear rate.
This implies that $\sigma_{xy}(\bm{\Gamma})$ also scales as
$\epsilon^{1/2}$.
To illuminate this feature, we introduce the following scaling for the
anisotropic quantities,
\begin{eqnarray}
\bm{X}^{T} \cdot \bm{\Sigma}_{xy} \cdot \bm{X}
=
\epsilon^{1/2}
\tilde{\bm{X}}^{T} \cdot \bm{\Sigma}_{xy} \cdot \tilde{\bm{X}},
\end{eqnarray}
where $\bm{X}$ is an arbitrary anisotropic vector and
$(\bm{\Sigma}_{xy})_{\mu\nu}=\delta_{\mu x}\delta_{\nu y}$.
For instance, we have
$
p_{i,x}^{*} p_{i,y}^{*}
=
\epsilon^{1/2}
\tilde{p}_{i,x} \tilde{p}_{i,y}.
$

Then, we can expand $i\mathcal{L}^{\dagger}(\bm{\Gamma})$ in terms of
$\epsilon$ as
\begin{eqnarray}
i\mathcal{L}^{*\dagger}(\bm{\Gamma})
=
i\mathcal{L}^{(\mathrm{eq})*}(\bm{\Gamma})
+
\epsilon
i\tilde{\mathcal{L}}_{1}(\bm{\Gamma}),
\label{eq:Ldagger}
\end{eqnarray}
where the unperturbed operator
$i\mathcal{L}^{(\mathrm{eq})*}(\bm{\Gamma})$ is given by
Eq.~(\ref{eq:iLel}) and the perturbed operator
$i\tilde{\mathcal{L}}_{1}(\bm{\Gamma})$ reads
\begin{eqnarray}
i\tilde{\mathcal{L}}_{1}(\bm{\Gamma}) 
&=&
i\tilde{\mathcal{L}}_{\dot\gamma}(\bm{\Gamma}) 
+
i\tilde{\mathcal{L}}^{(\mathrm{vis})}(\bm{\Gamma})
+
\tilde{\Lambda}(\bm{\Gamma})
\end{eqnarray}
with
\begin{eqnarray}
i\tilde{\mathcal{L}}_{\dot\gamma}(\bm{\Gamma}) 
&=&
\tilde{\dot\gamma}
\sum_{i=1}^{N}
\left[
\tilde{y}_{i} \frac{\partial}{\partial \tilde{x}_{i}} 
-
\tilde{p}_{i,y} \frac{\partial}{\partial \tilde{p}_{i,x}} \,
\right],
\\
i\tilde{\mathcal{L}}^{(\mathrm{vis})}(\bm{\Gamma})
&=&
\sum_{i=1}^{N}
\tilde{\bm{F}}_{i}^{(\mathrm{vis})}
\cdot
\frac{\partial}{\partial \bm{p}^{*}_{i}},
\\
\tilde{\bm{F}}_{i}^{(\mathrm{vis})}
&=&
-\sum_{j\neq i}
\Theta(1-r^{*}_{ij})
(\dot{\bm{r}}^{*}_{ij}\cdot\hat{\bm{r}}_{ij})
\hat{\bm{r}}_{ij},
\\
\tilde{\Lambda}(\bm{\Gamma})
&=&
-\sum_{i,j}{}^{'}
\Theta(1-r^{*}_{ij}),
\end{eqnarray}
respectively. 
%
%
%\vspace{1em}
%
Accordingly, we expand the distribution function and the eigenvalue as
\begin{eqnarray}
\Psi_{n}^{*}(\bm{\Gamma})
&=&
\rho_{\mathrm{eq}}^{*}(\bm{\Gamma})
\left[ \Psi_{n}^{(0)*}(\bm{\Gamma}) 
+ 
\epsilon \tilde{\Psi}_{n}^{(1)}(\bm{\Gamma})
\right]
+
\mathcal{O}(\epsilon^2),
\hspace{1em}
\label{eq:Psi}
\\
z_{n}^{*}
&=&
z_{n}^{(0)*}
+
\epsilon \tilde{z}_{n}^{(1)}
+
\mathcal{O}(\epsilon^2),
\label{eq:omega}
\end{eqnarray}
where $\rho_{\mathrm{eq}}^{*}(\bm{\Gamma})$ is the canonical
distribution.
Note that $\rho_{\mathrm{eq}}^{*}(\bm{\Gamma})$ satisfies
\begin{eqnarray}
i\mathcal{L}^{(\mathrm{eq})*}(\bm{\Gamma})
\rho_{\mathrm{eq}}^{*}(\bm{\Gamma})
=0 
\end{eqnarray}
by virtue of the conservation of the internal energy.
%,
%%
%\begin{eqnarray}
%&&
%\hspace{-2em}
%i\mathcal{L}_{\mathrm{eq}}(\bm{\Gamma}) 
%H_{\mathrm{eq}}(\bm{\Gamma})
%=
%\sum_{i=1}^{N}
%\left[
%\tilde{\bm{p}}_{i}
%\cdot
%\frac{\partial}{\partial \tilde{\bm{r}}_{i}}
%+
%\tilde{\bm{F}}_{i}^{(\mathrm{el})}
%\cdot
%\frac{\partial}{\partial \tilde{\bm{p}}_{i}}
%\right]
%H_{\mathrm{eq}}(\bm{\Gamma})
%\nonumber \\
%%
%&=&
%\sum_{i=1}^{N}
%\left[
%\tilde{\bm{p}}_{i}
%\cdot
%\frac{\partial U}{\partial \tilde{\bm{r}}_{i}}
%+
%\tilde{\bm{F}}_{i}^{(\mathrm{el})}
%\cdot
%\tilde{\bm{p}}_{i}
%\right]
%=
%0.
%\end{eqnarray}
%%
From Eqs.~(\ref{eq:LiouvilleEq}), (\ref{eq:Ldagger}), (\ref{eq:Psi}),
and (\ref{eq:omega}), we obtain
%
%\begin{widetext}
%
\begin{eqnarray}
&&
\hspace{-2em}
i\mathcal{L}^{(\mathrm{eq})*}(\bm{\Gamma})
\Psi_{n}^{(0)*}(\bm{\Gamma})
=
-z_{n}^{(0)*}
\Psi_{n}^{(0)*}(\bm{\Gamma}),
\label{eq:order0}
\\
&&
\hspace{-2em}
i\mathcal{L}^{(\mathrm{eq})*}(\bm{\Gamma})
\tilde{\Psi}_{n}^{(1)}(\bm{\Gamma})
+
\rho_{\mathrm{eq}}^{*}(\bm{\Gamma})^{-1}
i\tilde{\mathcal{L}}_{1}(\bm{\Gamma})
\rho_{\mathrm{eq}}^{*}(\bm{\Gamma})
\Psi_{n}^{(0)*}(\bm{\Gamma})
\nonumber \\
&=&
-z_{n}^{(0)*}
\tilde{\Psi}_{n}^{(1)}(\bm{\Gamma})
-
\tilde{z}_{n}^{(1)}
\Psi_{n}^{(0)*}(\bm{\Gamma}).
\label{eq:order1}
\end{eqnarray}
%
%\end{widetext}
%
%\vspace{1em}
%

%\vspace{1em}
\subsection{Eigenequation for the zero modes}

The unperturbed operator $i\mathcal{L}^{(\mathrm{eq})*}(\bm{\Gamma})$
has degenerate five zero-modes, which we denote by
$\phi_{\alpha}^{*}(\bm{\Gamma})$ ($\alpha=1,\cdots,5$),
\begin{eqnarray}
i\mathcal{L}^{(\mathrm{eq})*}(\bm{\Gamma}) 
\phi_{\alpha}^{*}(\bm{\Gamma})
=
0
\hspace{1em}
(\alpha = 1,\cdots,5).
\end{eqnarray}
Explicitly, they are given by
\begin{eqnarray}
\phi_{\alpha}^{*}(\bm{\Gamma})
&\propto&
\left\{
1, \,
\sum_{i=1}^{N} p_{i,x}^{*}, \,
\sum_{i=1}^{N} p_{i,y}^{*}, \,
\sum_{i=1}^{N} p_{i,z}^{*}, \,
\mathcal{H}^{*}(\bm{\Gamma})
\right\}.
\hspace{2em}
\label{eq:func_basis}
\end{eqnarray}
The equalities 
\begin{eqnarray}
i\mathcal{L}^{(\mathrm{eq})*}(\bm{\Gamma}) 
\sum_{i=1}^{N}
p_{i,\mu}^{*}
=
\sum_{i=1}^{N}
F_{i,\mu}^{(\mathrm{el})*}
=0 \,\,
(\mu = x,y,z)
\hspace{1em}
\end{eqnarray}
follow from the conservation of the momentum.
%
%
%\vspace{1em}
%
We choose $\{ \phi_{\alpha}^{*}(\bm{\Gamma}) \}$ ($\alpha=1,\cdots,5$)
to be orthogonal, i.e.
\begin{eqnarray}
\int d\bm{\Gamma}^{*}\, 
\rho_{\mathrm{eq}}^{*}(\bm{\Gamma})
\phi_{\alpha}^{*}(\bm{\Gamma}) \phi_{\alpha'}^{*}(\bm{\Gamma})
\propto
\delta_{\alpha\alpha'}.
\end{eqnarray}
Specifically, we adopt
\begin{eqnarray}
\phi_{1}^{*}(\bm{\Gamma})
&=&
1,
\label{eq:phi1}
\\
\phi_{2}^{*}(\bm{\Gamma})
&=&
\frac{1}{\sqrt{\frac{3}{2}N}T^{*}}
\left(
\sum_{i=1}^{N}
\frac{\bm{p}_{i}^{*2}}{2}
-
\frac{3}{2}NT^{*}
\right),
\label{eq:phi2}
\\
\phi_{\alpha}^{*}(\bm{\Gamma})
&=&
\frac{1}{\sqrt{NT^{*}}}
\sum_{i=1}^{N} p_{i,\lambda}^{*}
\label{eq:phi_345}
\end{eqnarray}
where $\lambda=x$, $y$, and $z$ correspond to $\alpha=3$, 4, and 5,
respectively. 
This leads to
\begin{eqnarray}
\int d\bm{\Gamma}^{*}\, 
\rho_{\mathrm{eq}}^{*}(\bm{\Gamma})
\phi_{\alpha}^{*}(\bm{\Gamma}) \phi_{\alpha'}^{*}(\bm{\Gamma})
=
\delta_{\alpha\alpha'}.
\end{eqnarray}
%
%
%[unofficial comment: note that $\phi_{\alpha}(\bm{\Gamma})$
%($\alpha=2,\cdots,5$) scale as $N^{1/2}$ and diverge in the
%thermodynamic limit.]
%
As for the energy eigenmode, $\phi_{2}^{*}(\bm{\Gamma})$, we consider
only the kinetic energy.
We expect this treatment to be valid for the unperturbed zero-modes in
the hard-core limit.
%
%This is consistent with our treatment where we take the hard-core limit
%later.
%

In the following, we work in the 5-dimensional space spanned by
Eqs.~(\ref{eq:phi1})--(\ref{eq:phi_345}). 
Then, the distribution function is explicitly given by
\begin{eqnarray}
\rho^{*}(\bm{\Gamma},t)
&=&
\sum_{\alpha =1}^{5}
e^{\epsilon \tilde{z}_{\alpha}^{(1)}t^{*}+\cdots} 
\rho_{\mathrm{eq}}^{*}(\bm{\Gamma})
\nonumber \\
&&
\times
\left[
\Psi_{\alpha}^{(0)*}(\bm{\Gamma}) 
+
\epsilon \tilde{\Psi}_{\alpha}^{(1)}(\bm{\Gamma})  
+
\cdots
\right].
\end{eqnarray}
%
%
%\vspace{1em}
%
Since $\{ \phi_{\alpha}^{*}(\bm{\Gamma}) \}$ ($\alpha=1,\cdots,5$) are
five-fold degenerate, we must choose an appropriate linear combination
to construct the unperturbed distribution function,
\begin{eqnarray}
\Psi_{\alpha}^{(0)*}(\bm{\Gamma})
=
\sum_{\alpha'=1} ^{5}
c_{\alpha\alpha'}
\phi_{\alpha'}^{*}(\bm{\Gamma}),
\label{eq:Psi0}
\end{eqnarray}
where $\{ c_{\alpha\beta}\}_{\alpha,\beta=1}^{5}$ are the coefficients.

%\vspace{1em}
%
We next consider Eq.~(\ref{eq:order1}).
% for the zero modes, i.e. $n=\alpha=1,\cdots, 5$. 
%
Since $z_{\alpha}^{(0)*}=0$, we obtain
\begin{eqnarray}
&&
\hspace{-2em}
\rho_{\mathrm{eq}}^{*}(\bm{\Gamma})
i\mathcal{L}^{(\mathrm{eq})*}(\bm{\Gamma})
\tilde{\Psi}_{\alpha}^{(1)}(\bm{\Gamma})
+
i\tilde{\mathcal{L}}_{1}(\bm{\Gamma})
\rho_{\mathrm{eq}}^{*}(\bm{\Gamma})
\Psi_{\alpha}^{(0)*}(\bm{\Gamma})
\nonumber \\
&=&
-\tilde{z}_{\alpha}^{(1)}
\rho_{\mathrm{eq}}^{*}(\bm{\Gamma})
\Psi_{\alpha}^{(0)*}(\bm{\Gamma}). 
\label{eq:order1_zeromode}
\end{eqnarray}
By multiplying Eq.~(\ref{eq:order1_zeromode}) by
$\phi_{\alpha'}^{*}(\bm{\Gamma})$ and integrating with respect to
$\bm{\Gamma}^{*}$, we obtain
\begin{eqnarray}
&&
\hspace{-2em}
\int d\bm{\Gamma}^{*} \,
\phi_{\alpha'}^{*}(\bm{\Gamma})
\left[
i\tilde{\mathcal{L}}_{1}(\bm{\Gamma})
\rho_{\mathrm{eq}}^{*}(\bm{\Gamma})
\Psi_{\alpha}^{(0)*}(\bm{\Gamma})
\right]
\nonumber \\
&=&
-\tilde{z}_{\alpha}^{(1)}
\int d\bm{\Gamma}^{*} \,
\phi_{\alpha'}^{*}(\bm{\Gamma})
\left[
\rho_{\mathrm{eq}}^{*}(\bm{\Gamma})
\Psi_{\alpha}^{(0)*}(\bm{\Gamma})
\right],
\label{eq:order1_eigen}
\end{eqnarray}
which follows from the fact that the first term on the left-hand side
of Eq.~(\ref{eq:order1_zeromode}) vanishes,
\begin{eqnarray}
\int d\bm{\Gamma}^{*} \,
\phi_{\alpha'}^{*}(\bm{\Gamma})
\left[
\rho_{\mathrm{eq}}^{*}(\bm{\Gamma})
i\mathcal{L}^{(\mathrm{eq})*}(\bm{\Gamma})
\tilde{\Psi}_{\alpha}^{(1)}(\bm{\Gamma}) 
\right]
&=&
0,
\end{eqnarray}
due to $i\mathcal{L}^{(\mathrm{eq})*} \phi_{\alpha}^{*}(\bm{\Gamma}) =
0$.
From Eq.~(\ref{eq:Psi0}), Eq.~(\ref{eq:order1_eigen}) is expressed as
\begin{eqnarray}
-\tilde{z}_{\alpha}^{(1)} 
c_{\alpha\alpha'}
=
\sum_{\alpha''}
c_{\alpha\alpha''}
W_{\alpha'\alpha''},
\label{eq:order1_eigen2}
\end{eqnarray}
where
\begin{eqnarray}
W_{\alpha'\alpha''}
\equiv 
\int d\bm{\Gamma}^{*}\,
\phi_{\alpha'}^{*}(\bm{\Gamma})
i\tilde{\mathcal{L}}_{1}(\bm{\Gamma}) 
\rho_{\mathrm{eq}}^{*}(\bm{\Gamma}) 
\phi_{\alpha''}^{*}(\bm{\Gamma}).
\hspace{2em}
\end{eqnarray}
We obtain five eigenvalue equations for $\tilde{z}_{\alpha}^{(1)}$
($\alpha=1,\cdots,5$) from Eq.~(\ref{eq:order1_eigen2}),
\begin{eqnarray}
\bm{W} \bm{c}_{\alpha}^{T} 
=
-\tilde{z}_{\alpha}^{(1)}
\bm{c}_{\alpha}^{T},
\label{eq:EigenEq_vector}
\end{eqnarray}
where $\bm{c}_{\alpha}^{T}$ is a vector which constitutes a matrix
$\bm{c}^{T}$, i.e. $\bm{c}^{T} = \{ \bm{c}_{1}^{T}, \cdots,
\bm{c}_{5}^{T}\}$.
Here, the superscript $T$ denotes a transpose.
Then, the eigen equations read
\begin{eqnarray}
\det 
\left[ \bm{W} + \tilde{z}_{\alpha}^{(1)} \bm{1}\right] 
=
0
\hspace{1em}
(\alpha = 1,\cdots,5),
\label{eq:EigenEq}
\end{eqnarray}
where $\bm{1}$ is the identity matrix.

%\vspace{1em}
%
Now we calculate the matrix elements of $\bm{W}$, which can be
decomposed into
\begin{eqnarray}
\bm{W}
=
\bm{W}^{(\dot\gamma)} 
+
\bm{W}^{(\mathrm{vis})}
+
\bm{W}^{(\Lambda)},
\end{eqnarray}
where
\begin{eqnarray}
W_{\alpha\alpha'}^{(\dot\gamma)}
&=&
\int d\bm{\Gamma}^{*}\,
\phi_{\alpha}^{*}(\bm{\Gamma})
i\tilde{\mathcal{L}}_{\dot\gamma}(\bm{\Gamma})
\rho_{\mathrm{eq}}^{*}(\bm{\Gamma}) \phi_{\alpha'}^{*}(\bm{\Gamma}),
\label{eq:Wdotgamma}
\\
W_{\alpha\alpha'}^{(\mathrm{vis})}
&=&
\int d\bm{\Gamma}^{*}\,
\phi_{\alpha}^{*}(\bm{\Gamma})
i\tilde{\mathcal{L}}^{(\mathrm{vis})}(\bm{\Gamma})
\rho_{\mathrm{eq}}^{*}(\bm{\Gamma}) \phi_{\alpha'}^{*}(\bm{\Gamma}),
\hspace{1em}
\label{eq:Wvis}
\\
W_{\alpha\alpha'}^{(\Lambda)}
&=&
\int d\bm{\Gamma}^{*}\,
\phi_{\alpha}^{*}(\bm{\Gamma})
\tilde{\Lambda}(\bm{\Gamma})
\rho_{\mathrm{eq}}^{*}(\bm{\Gamma}) \phi_{\alpha'}^{*}(\bm{\Gamma}).
\label{eq:WLambda}
\end{eqnarray}
For Eqs.~(\ref{eq:Wdotgamma}) and (\ref{eq:Wvis}), it is necessary to
evaluate the integrands,
\begin{eqnarray}
&&
\hspace{-2em}
i\tilde{\mathcal{L}}_{\dot\gamma}(\bm{\Gamma}) 
\rho_{\mathrm{eq}}^{*}(\bm{\Gamma})
\phi_{\alpha}^{*}(\bm{\Gamma})
\nonumber \\
&=&
\left[
i\tilde{\mathcal{L}}_{\dot\gamma}(\bm{\Gamma}) 
\rho_{\mathrm{eq}}^{*}(\bm{\Gamma})
\right]
\phi_{\alpha}^{*}(\bm{\Gamma})
+
\rho_{\mathrm{eq}}^{*}(\bm{\Gamma})
i\tilde{\mathcal{L}}_{\dot\gamma}(\bm{\Gamma}) 
\phi_{\alpha}^{*}(\bm{\Gamma}),
\nonumber \\
\\
&&
\hspace{-2em}
i\tilde{\mathcal{L}}^{(\mathrm{vis})}(\bm{\Gamma}) 
\rho_{\mathrm{eq}}^{*}(\bm{\Gamma})
\phi_{\alpha}^{*}(\bm{\Gamma})
\nonumber \\
&=&
\left[
i\tilde{\mathcal{L}}^{(\mathrm{vis})}(\bm{\Gamma}) 
\rho_{\mathrm{eq}}^{*}(\bm{\Gamma})
\right]
\phi_{\alpha}^{*}(\bm{\Gamma})
+
\rho_{\mathrm{eq}}^{*}(\bm{\Gamma})
i\tilde{\mathcal{L}}^{(\mathrm{vis})}(\bm{\Gamma}) 
\phi_{\alpha}^{*}(\bm{\Gamma}).
\nonumber \\
\end{eqnarray}
The Liouvillians act on $\rho_{\mathrm{eq}}^{*}(\bm{\Gamma})$ as
\begin{eqnarray}
&&
\hspace{-2em}
\rho_{\mathrm{eq}}^{*}(\bm{\Gamma})^{-1}
i\tilde{\mathcal{L}}_{\dot\gamma}(\bm{\Gamma}) 
\rho_{\mathrm{eq}}^{*}(\bm{\Gamma})
\nonumber \\
&=&
-\beta^{*}\tilde{\dot\gamma}
\sum_{i=1}^{N}
\left[
\tilde{y}_{i}
\frac{\partial}{\partial \tilde{x}}_{i}
-
\tilde{p}_{i,y}
\frac{\partial}{\partial \tilde{p}_{i,x}}
\right]
\mathcal{H}^{*}(\bm{\Gamma})
\nonumber \\
&=&
\beta^{*}\tilde{\dot\gamma}
\sum_{i=1}^{N}
\left[
\tilde{p}_{i,x} \tilde{p}_{i,y}
+
\tilde{y}_{i} \tilde{F}_{i,x}^{(\mathrm{el})}
\right]
%\nonumber \\
%%
%&=&
=
\beta^{*}
\tilde{\dot\gamma} V^{*}
\tilde{\sigma}_{xy}^{(\mathrm{el})}(\bm{\Gamma}),
\\
&&
\hspace{-2em}
\rho_{\mathrm{eq}}^{*}(\bm{\Gamma})^{-1}
i\tilde{\mathcal{L}}^{(\mathrm{vis})}(\bm{\Gamma}) 
\rho_{\mathrm{eq}}^{*}(\bm{\Gamma})
\nonumber \\
&=&
-\beta^{*}
\sum_{i=1}^{N}
\tilde{\bm{F}}_{i}^{(\mathrm{vis})}
\cdot
\frac{\partial}{\partial \bm{p}^{*}_{i}}
\mathcal{H}^{*}(\bm{\Gamma})
=
-\beta^{*}
\sum_{i=1}^{N}
\bm{p}^{*}_{i}
\cdot
\tilde{\bm{F}}_{i}^{(\mathrm{vis})}
\nonumber \\
&=&
\frac{1}{2}\beta^{*}
\sum_{i,j}{}^{'}
\Theta(1-r^{*}_{ij})
\left( \dot{\bm{r}}^{*}_{ij}\cdot\hat{\bm{r}}_{ij}\right)
\left(\bm{p}^{*}_{ij}\cdot\hat{\bm{r}}_{ij}\right)
\nonumber \\
&=&
\frac{1}{2}\beta^{*}
\sum_{i,j}{}^{'}
\Theta(1-r^{*}_{ij})
\left( \left[ \bm{p}^{*}_{ij} + \tilde{\dot\gamma} \tilde{y}_{i} 
\bm{e}_{x} \right]\cdot\hat{\bm{r}}_{ij}\right)
\left(\bm{p}^{*}_{ij}\cdot\hat{\bm{r}}_{ij}\right)
\nonumber \\
&=&
2\beta^{*}
\tilde{\mathcal{R}}^{(1)}(\bm{\Gamma})
-
\epsilon^{1/2}\beta^{*}\tilde{\dot\gamma}V^{*}
\tilde{\sigma}_{xy}^{(\mathrm{vis})(1)}(\bm{\Gamma})
\nonumber \\
&\approx&
2\beta^{*}
\tilde{\mathcal{R}}^{(1)}(\bm{\Gamma}),
\end{eqnarray}
where the scaled quantities
$\tilde{\sigma}_{xy}^{(\mathrm{el})}(\bm{\Gamma})$,
$\tilde{\sigma}_{xy}^{(\mathrm{vis})(1)}(\bm{\Gamma})$, and
$\tilde{\mathcal{R}}^{(1)}(\bm{\Gamma})$ are given by
\begin{eqnarray}
\tilde{\sigma}_{xy}^{(\mathrm{el})}(\bm{\Gamma})
&=&
\frac{1}{V^*} 
\sum_{i=1}^{N}
\left[
\tilde{p}_{i,x} \tilde{p}_{i,y}
+
\tilde{y}_i \tilde{F}_{i,x}^{(\mathrm{el})}
\right], 
\\
\tilde{\sigma}_{xy}^{\rm (vis)(1)}(\bm{\Gamma}) 
&=&
-\frac{1}{2V^*}
\sum_{i,j}{}^{'} 
\left( \bm{p}_{ij}^{*} \cdot \hat{\bv{r}}_{ij} \right)
\tilde{\hat{x}}_{ij} \tilde{\hat{y}}_{ij} r_{ij}^{*} 
\Theta(1-r_{ij}^{*}),
\hspace{2.5em}
\\
\tilde{\mathcal{R}}^{(1)}(\bm{\Gamma})
&=&
\frac{1}{4}
\sum_{i,j}{}^{'} 
\left( \bm{p}_{ij}^{*} \cdot \hat{\bv{r}}_{ij} \right)^{2} 
\Theta(1-r_{ij}^{*}).
\end{eqnarray}
On the other hand, the Liouvillians act on
$\phi_{\alpha}^{*}(\bm{\Gamma})$ as
\begin{eqnarray}
&&
\hspace{-2em}
i\tilde{\mathcal{L}}_{\dot\gamma}(\bm{\Gamma}) 
\phi_{\alpha}^{*}(\bm{\Gamma})
=
0
\hspace{1em} (\alpha = 1,4,5),
\label{eq:iLdotgamma_phi1}
\\
&&
\hspace{-2em}
i\tilde{\mathcal{L}}_{\dot\gamma}(\bm{\Gamma}) 
\phi_{2}^{*}(\bm{\Gamma})
=
%\dot\gamma
%\sum_{i=1}^{N}
%\left[
%y_{i}
%\frac{\partial}{\partial x_{i}}
%-
%p_{i}^{y}
%\frac{\partial}{\partial p_{i}^{x}}
%\right]
%\frac{1}{N\sqrt{\frac{3}{2}}T}
%\nonumber \\
%%
%&&
%\times
%\left(
%\sum_{j=1}^{N}
%\frac{\bm{p}_{j}^2}{2}
%-
%\frac{3}{2}
%NT
%\right)
%=
-\frac{\tilde{\dot\gamma}}{\sqrt{\frac{3}{2}N} T^{*}}
\sum_{i=1}^{N}
\tilde{p}_{i,x} \tilde{p}_{i,y},
\\
&&
\hspace{-2em}
i\tilde{\mathcal{L}}_{\dot\gamma}(\bm{\Gamma}) 
\phi_{3}^{*}(\bm{\Gamma})
%=
%\dot\gamma
%\sum_{i=1}^{N}
%\left[
%y_{i}
%\frac{\partial}{\partial x_{i}}
%-
%p_{i}^{y}
%\frac{\partial}{\partial p_{i}^{x}}
%\right]
%\frac{1}{N\sqrt{T}}
%\sum_{j=1}^{N} p_{j}^x
%\nonumber \\
%%
%&=&
=
-\frac{\tilde{\dot\gamma}}{\sqrt{NT^{*}}}
\sum_{i=1}^{N}
\tilde{p}_{i,y},
\label{eq:iLdotgamma_phi5}
\end{eqnarray}
and
\begin{eqnarray}
&&
%\hspace{-2em}
i\tilde{\mathcal{L}}^{(\mathrm{vis})}(\bm{\Gamma}) 
\phi_{1}^{*}(\bm{\Gamma})
=
0,
\label{eq:iLvis_phi1}
\\
&&
%\hspace{-2em}
i\tilde{\mathcal{L}}^{(\mathrm{vis})}(\bm{\Gamma}) 
\phi_{2}^{*}(\bm{\Gamma})
%=
%\sum_{i=1}^{N}
%\bm{F}_{i}^{(\mathrm{vis})}
%\cdot
%\frac{\partial}{\partial \bm{p}_{i}}
%\frac{1}{N\sqrt{\frac{3}{2}}T}
%\nonumber \\
%%
%&&
%\times
%\left(
%\sum_{j=1}^{N}
%\frac{\bm{p}_{j}^2}{2}
%-
%\frac{3}{2}
%NT
%\right)
=
\frac{1}{\sqrt{\frac{3}{2}N} T^{*}}
\sum_{i=1}^{N}
\bm{p}^{*}_{i}
\cdot\tilde{\bm{F}}_{i}^{(\mathrm{vis})}
%\nonumber \\
%%
%&=&
=
\frac{-2\tilde{\mathcal{R}}^{(1)}(\bm{\Gamma})}{\sqrt{\frac{3}{2}N}
T^{*}},
\nonumber \\
\\
&&
%\hspace{-2em}
i\tilde{\mathcal{L}}^{(\mathrm{vis})}(\bm{\Gamma}) 
\phi_{\alpha}^{*}(\bm{\Gamma})
%=
%\sum_{i=1}^{N}
%\bm{F}_{i}^{(\mathrm{vis})}
%\cdot
%\frac{\partial}{\partial \bm{p}_{i}}
%\frac{1}{N\sqrt{T}}
%\sum_{j=1}^{N} p_{j}^x
%\nonumber \\
%%
%&=&
=
\sum_{i=1}^{N}
\frac{\tilde{F}_{i,\lambda}^{(\mathrm{vis})}}{\sqrt{NT^{*}}}
\hspace{1em} (\alpha=3,4,5),
\label{eq:iLvis_phi5}
\end{eqnarray}
where $\lambda=x$, $y$, and $z$ correspond to $\alpha=3$, 4, and 5,
respectively.
Hence, Eqs.~(\ref{eq:Wdotgamma}) and (\ref{eq:Wvis}) are recasted in the
form
%
%\begin{widetext}
%
\begin{eqnarray}
&&
\hspace{-2em}
W_{\alpha\alpha'}^{(\dot\gamma)}
=
\int d\bm{\Gamma}^{*}\,
\rho_{\mathrm{eq}}^{*}(\bm{\Gamma})
\phi_{\alpha}^{*}(\bm{\Gamma})
\beta^{*}\tilde{\dot\gamma} V^{*}
\tilde{\sigma}_{xy}^{(\mathrm{el})}(\bm{\Gamma})
\phi_{\alpha'}^{*}(\bm{\Gamma})
\nonumber \\
&&
+
\int d\bm{\Gamma}^{*}\,
\rho_{\mathrm{eq}}^{*}(\bm{\Gamma})
\phi_{\alpha}^{*}(\bm{\Gamma})
i\tilde{\mathcal{L}}_{\dot\gamma}(\bm{\Gamma})
\phi_{\alpha'}^{*}(\bm{\Gamma}),
\\
&&
\hspace{-2em}
W_{\alpha\alpha'}^{(\mathrm{vis})}
=
\int d\bm{\Gamma}^{*}\,
\rho_{\mathrm{eq}}^{*}(\bm{\Gamma})
\phi_{\alpha}^{*}(\bm{\Gamma})
2\beta^{*}
\tilde{\mathcal{R}}^{(1)}(\bm{\Gamma})
\phi_{\alpha'}^{*}(\bm{\Gamma})
\nonumber \\
&&
+
\int d\bm{\Gamma}^{*}\,
\rho_{\mathrm{eq}}^{*}(\bm{\Gamma})
\phi_{\alpha}^{*}(\bm{\Gamma})
i\tilde{\mathcal{L}}^{(\mathrm{vis})}(\bm{\Gamma})
\phi_{\alpha'}^{*}(\bm{\Gamma}).
\end{eqnarray}
%
%\end{widetext}
%
Together with Eq.~(\ref{eq:WLambda}), we obtain
%
%\begin{widetext}
%
\begin{eqnarray}
&&
\hspace{-2em}
W_{\alpha\alpha'} 
=
\int d\bm{\Gamma}^{*}\,
\rho_{\mathrm{eq}}^{*}(\bm{\Gamma})
\phi_{\alpha}^{*}(\bm{\Gamma})
\left[
\beta^{*}\tilde{\dot\gamma} V^{*}
\tilde{\sigma}_{xy}^{(\mathrm{el})}(\bm{\Gamma})
\right.
\nonumber \\
&&
\left.
+
2\beta^{*}\Delta\tilde{\mathcal{R}}^{(1)}(\bm{\Gamma})
+
i\tilde{\mathcal{L}}_{\dot\gamma}(\bm{\Gamma})
+
i\tilde{\mathcal{L}}^{(\mathrm{vis})}(\bm{\Gamma})
\right]
\phi_{\alpha'}^{*}(\bm{\Gamma}),
\hspace{1em}
\end{eqnarray}
%
%\end{widetext}
%
where $\Delta\tilde{\mathcal{R}}^{(1)}(\bm{\Gamma})$ is given by
\begin{eqnarray}
\Delta\tilde{\mathcal{R}}^{(1)}(\bm{\Gamma})
&=&
\tilde{\mathcal{R}}^{(1)}(\bm{\Gamma})
+
\frac{T^{*}}{2}
\tilde{\Lambda}(\bm{\Gamma}).
\end{eqnarray}
%

%\vspace{1em}
%
We denote the three components of $\bm{W}$ as
\begin{eqnarray}
\bm{W}
=
\bm{W}^{(\mathrm{s})} 
+
\bm{W}^{(\mathrm{a1})}
+
\bm{W}^{(\mathrm{a2})},
\label{eq:W_decomp}
\end{eqnarray}
where
\begin{eqnarray}
&&
\hspace{-2em}
W_{\alpha\alpha'}^{(\mathrm{s})}
\equiv
\int d\bm{\Gamma}^{*}\,
\rho_{\mathrm{eq}}^{*}(\bm{\Gamma})
\phi_{\alpha}^{*}(\bm{\Gamma})
\nonumber \\
&&
\times
\beta^{*}
\left[
\tilde{\dot\gamma} V^{*}
\tilde{\sigma}_{xy}^{(\mathrm{el})}
+
2\Delta\tilde{\mathcal{R}}^{(1)} \right]
\phi_{\alpha'}^{*}(\bm{\Gamma})
\end{eqnarray}
is the symmetric part and 
\begin{eqnarray}
W_{\alpha\alpha'}^{(\mathrm{a1})}
&\equiv&
\int d\bm{\Gamma}^{*}\,
\rho_{\mathrm{eq}}^{*}(\bm{\Gamma})
\phi_{\alpha}^{*}(\bm{\Gamma})
i\tilde{\mathcal{L}}_{\dot\gamma}(\bm{\Gamma})
\phi_{\alpha'}^{*}(\bm{\Gamma}),
\\
W_{\alpha\alpha'}^{(\mathrm{a2})}
&\equiv&
\int d\bm{\Gamma}^{*}\,
\rho_{\mathrm{eq}}^{*}(\bm{\Gamma})
\phi_{\alpha}^{*}(\bm{\Gamma})
i\tilde{\mathcal{L}}^{(\mathrm{vis})}(\bm{\Gamma})
\phi_{\alpha'}^{*}(\bm{\Gamma}),
\end{eqnarray}
are the asymmetric parts.
%
%\vspace{1em}
%
The elements of $\bm{W}^{(\mathrm{a1})}$ and $\bm{W}^{(\mathrm{a2})}$
are given from
Eqs.~(\ref{eq:iLdotgamma_phi1})--(\ref{eq:iLdotgamma_phi5}) and
Eqs.~(\ref{eq:iLvis_phi1})--(\ref{eq:iLvis_phi5}) as 
%
%%%%%%%%%%%%%%%%%%
\begin{comment}
%%%%%%%%%%%%%%%%%%
\begin{eqnarray}
W_{\alpha, 1}^{(\mathrm{a1})} 
&=&
0
\hspace{1em} (\alpha = 1,\cdots,5),
\\
%
W_{\alpha, 2}^{(\mathrm{a1})} 
&=&
0
\hspace{1em} (\alpha = 1,\cdots,5),
\\
%
W_{4, 3}^{(\mathrm{a1})} 
&=&
-\frac{\dot\gamma}{NmT}
\int d\bm{\Gamma} \,
\rho_{\mathrm{eq}}(\bm{\Gamma})
\sum_{i,j=1}^{N}
p_{i}^y p_{j}^y
= 
-\dot\gamma,
\hspace{1em}
W_{\alpha, 3}^{(\mathrm{a1})} 
=
0
\hspace{1em} (\alpha \neq 4),
\\
%
W_{\alpha, 4}^{(\mathrm{a1})} 
&=&
0
\hspace{1em} (\alpha = 1,\cdots,5),
\\
%
W_{\alpha, 5}^{(\mathrm{a1})} 
&=&
0
\hspace{1em} (\alpha = 1,\cdots,5),
\end{eqnarray}
%
i.e.
%%%%%%%%%%%%%%%%%%
\end{comment}
%%%%%%%%%%%%%%%%%%
%
%\begin{widetext}
%
\begin{eqnarray}
\bm{W}^{(\mathrm{a1})} 
%&=&
\hspace{-0.2em}
&=&
\hspace{-0.2em}
\left[
\begin{array}{ccccc}
0 & 0 & 0 & 0 & 0 \\
0 & 0 & 0 & 0 & 0 \\
0 & 0 & 0 & 0 & 0 \\
0 & 0 & -\tilde{\dot\gamma} & 0 & 0 \\
0 & 0 & 0 & 0 & 0 
\end{array}
\right] \hspace{-0.2em},
\label{eq:Wa1}
\\
\hspace{0.5em}
\bm{W}^{(\mathrm{a2})} 
%&\approx&
\hspace{-0.2em}
&\approx&
\hspace{-0.2em}
\left[
\begin{array}{ccccc}
0 & -\sqrt{\frac{2}{3}N}\mathscr{G}
& 0 & 0 & 0 \\
0 & -\frac{2}{3}\mathscr{G} & 0 & 0 & 0 \\
0 & 0 & 0 & 0 & 0 \\
0 & 0 & 0 & 0 & 0 \\
0 & 0 & 0 & 0 & 0 
\end{array}
\right],
\label{eq:Wa2}
%\nonumber \\
\end{eqnarray}
where 
%$\bm{F}_{i}^{(\mathrm{vis})(1)}$,
%$\bm{F}_{i}^{(\mathrm{vis})(2)}$, 
$\mathscr{G}$
%, and $\mathscr{G}^{\mu\nu\lambda}(d)$ 
is given by
\begin{eqnarray}
%\bm{F}_{i}^{(\mathrm{vis})} 
%&=&
%\bm{F}_{i}^{(\mathrm{vis})(1)} 
%+
%\bm{F}_{i}^{(\mathrm{vis})(2)},
%\\
%%
%\bm{F}_{i}^{(\mathrm{vis})(1)} 
%&=&
%-\zeta
%\sum_{j\neq i}
%\Theta(d-r_{ij})
%\left( \frac{\bm{p}_{ij}}{m} \cdot \hat{\bm{r}}_{ij}\right)
%\hat{\bm{r}}_{ij}
%\\
%%
%\bm{F}_{i}^{(\mathrm{vis})(2)}
%&=&
%-\dot\gamma\zeta
%\sum_{j\neq i}
%\Theta(d-r_{ij})
%\hat{x}_{ij} \hat{y}_{ij}
%r_{ij}
%\hat{\bm{r}}_{ij},
%\\
%%
\mathscr{G}
&=&
n^{*} \int d^3\bm{r}^{*} \,
g(r^{*}, \varphi)
\Theta(1-r^{*}).
%\\
%%
%\mathscr{G}^{\mu\nu\lambda}(d)
%&=&
%\int d^3\bm{r} \,
%\Theta(r-d)
%g(\bm{r})\hat{r}^{\mu} \hat{r}^{\nu} \hat{r}^{\lambda}
%\frac{r}{d}.
\end{eqnarray}
%
%$\tilde{n}=nd^3$ is the non-dimesionalized average density.
%
Here, $g(r,\varphi)$ is the radial distribution function at equilibrium
defined by $\langle \sum_{i,j}{}' \,
\delta(\bm{r}-\bm{r}_{i})\rangle_{\mathrm{eq}} = N n g(r,\varphi)$.
In Eq.~(\ref{eq:Wa2}), terms of order $\epsilon^{1/2}$ are neglected.
%
%\vspace{1em}
%
%
The elements of $\bm{W}^{(s)}$ are given by
\begin{widetext}
\begin{eqnarray}
\bm{W}^{(\mathrm{s})} 
=
\left[
\begin{array}{ccccc}
0 & \sqrt{\frac{2}{3}N}\mathscr{G}
& 0 & 0 & 0 \\
\sqrt{\frac{2}{3}N}\mathscr{G}
& \frac{4}{3}\mathscr{G}
& 0 & 0 & 0 \\
0 & 0 & 2\mathscr{G}^{xx} & 
\tilde{\dot\gamma} + 2\mathscr{G}^{xy} &
2\mathscr{G}^{xz} \\
0 & 0 & 
\tilde{\dot\gamma} + 2\mathscr{G}^{yx}
& 
2\mathscr{G}^{yy}
& 
2\mathscr{G}^{yz} \\
0 & 0 & 
2\mathscr{G}^{zx}
& 
2\mathscr{G}^{zy}
& 
2\mathscr{G}^{zz}
\end{array}
\right]
=
\left[
\begin{array}{ccccc}
0 & \sqrt{\frac{2}{3}N}\mathscr{G}
& 0 & 0 & 0 \\
\sqrt{\frac{2}{3}N}\mathscr{G}
& \frac{4}{3}\mathscr{G}
& 0 & 0 & 0 \\
0 & 0 & \frac{2}{3}\mathscr{G} & \tilde{\dot\gamma} & 0 \\
0 & 0 & \tilde{\dot\gamma} & \frac{2}{3}\mathscr{G} & 0 \\
0 & 0 & 0 & 0 & \frac{2}{3}\mathscr{G}
\end{array}
\right]
\label{eq:Ws}
\end{eqnarray}
\end{widetext}
with
$
\mathscr{G}^{\mu\nu}
=
n^{*}
\int d^3\bm{r}^{*} \,
g(r^{*},\varphi)
\Theta(1-r^{*})
\hat{r}^{\mu}
\hat{r}^{\nu},
$
where the anisotropic components vanish, $\mathscr{G}^{\mu\nu}=0$
($\mu\neq\nu$), and the isotropic components are given by
$\mathscr{G}^{xx}$ = $\mathscr{G}^{yy}$ = $\mathscr{G}^{zz}$ =
$\mathscr{G}/3$.
From Eqs.~(\ref{eq:W_decomp}), (\ref{eq:Wa1}), (\ref{eq:Wa2}), and
(\ref{eq:Ws}), we obtain
%
%\begin{widetext}
%
\begin{eqnarray}
\bm{W}
=
\left[
\begin{array}{ccccc}
0 & 0 & 0 & 0 & 0 \\
\sqrt{\frac{2}{3}N} \mathscr{G}
& \frac{2}{3}\mathscr{G} & 0 & 0 & 0 \\
0 & 0 & \frac{2}{3}\mathscr{G} & \tilde{\dot\gamma} & 0 \\
0 & 0 & 0 & \frac{2}{3}\mathscr{G} & 0 \\
0 & 0 & 0 & 0 & \frac{2}{3}\mathscr{G}
\end{array}
\right].
\label{eq:W} 
\end{eqnarray}
%
%\end{widetext}
%
Then, from Eq.~(\ref{eq:EigenEq}), the eigenvalues
$\tilde{z}_{\alpha}^{(1)}$ ($\alpha=1,\cdots,5$) are obtained as
solutions of the following equation for $\lambda$,
\clearpage
\begin{widetext}
\begin{eqnarray}
\det
\left[
\begin{array}{ccccc}
\lambda & 0 & 0 & 0 & 0 \\
\sqrt{\frac{2}{3}N} \mathscr{G}
& \frac{2}{3}\mathscr{G} +\lambda
& 0 & 0 & 0 \\
0 & 0 & \frac{2}{3}\mathscr{G}+\lambda & 
\tilde{\dot\gamma} & 0 \\
0 & 0 & 0 & 
\frac{2}{3}\mathscr{G}+\lambda
& 0 \\
0 & 0 & 0 & 0 & 
\frac{2}{3}\mathscr{G}+\lambda 
\end{array}
\right]
=0.
\label{eq:EigenEq-2}
\end{eqnarray}
\end{widetext}

\subsection{Eigenvalues of the first order $z_{\alpha}^{(1)}$}

From Eq.~(\ref{eq:EigenEq-2}), the eigenvalues are given by
\begin{eqnarray}
\tilde{z}_{1}^{(1)} 
&=&
0,
\label{eq:z1_1}
\\
\tilde{z}_{\alpha}^{(1)} 
&=&
-\frac{2}{3}\mathscr{G}
\hspace{1em}
(\alpha = 2,3,4,5),
\label{eq:z2345_1}
\end{eqnarray}
from which we see that the four non-zero modes are degenerate and hence
can be approximated as a single relaxation mode.
The relaxation time for this mode is given by
\begin{eqnarray}
\tau^{*}_{\mathrm{rel}} 
\approx
-\frac{1}{\epsilon \tilde{z}_{\alpha}^{(1)}}
=
\left[
\frac{2}{3}\epsilon\, \mathscr{G}
\right]^{-1}.
\hspace{2em}
\label{eq:tau_rel_nondim}
\end{eqnarray}
%
%\end{widetext}
%
In the hard-core limit, $\mathscr{G}$ reduces to
\begin{eqnarray}
\mathscr{G}
\to
\sqrt{\pi}\,
\omega_{E}^{*}(T^{*}),
\label{eq:G1_hard-core}
\end{eqnarray}
(cf. Eq.~(\ref{eq:Theta_hard-core}) which will be shown later), where
$\omega_{E}(T)=4\sqrt{\pi}\,n\sqrt{T/m}\, g_{0}(\varphi)d^2$ is the
Enskog frequency of collisions, and $g_{0}(\varphi)$ is the first-peak
value of the radial distribution function,
i.e. $g_{0}(\varphi)=g(d,\varphi)$.
From Eqs.~(\ref{eq:tau_rel_nondim}) and (\ref{eq:G1_hard-core}), we
obtain
\begin{eqnarray}
\tau^{*}_{\mathrm{rel}} 
=
\left[ \frac{2\sqrt{\pi}}{3} \epsilon\,
\omega_{E}^{*}(T^{*})\right]^{-1}.
\label{eq:tau_rel_hard-core}
\end{eqnarray}

\section{Steady-state distribution function}
\label{subsec:SSD}

We derive an approximate explicit expression for the nonequilibrium
steady-state distribution function, with the aid of the relaxation time
derived in Sec~\ref{app:tauH}.
We start from an equilibrium state at $t\to -\infty$ and evolve the
system with shear and dissipation, reaching a nonequilibrium steady
state at $t=0$.
%
%%%%%
%
A formal but {\it exact} expression for the nonequilibrium steady-state
distribution function is given by~\cite{EM}
\begin{equation}
\rho_{\mathrm{SS}}^{(\mathrm{ex})}(\bm{\Gamma})
=
\exp
\left[
\int_{-\infty}^{0} d\tau \,
\Omega_{\mathrm{eq}}(\bm{\Gamma}(-\tau))
\right]
\rho_{\mathrm{eq}}(\bm{\Gamma}(-\infty)),
\label{app:eq:rho_SS_ex}
\end{equation}
which satisfies
$i\mathcal{L}^{\dagger}\rho_{\mathrm{SS}}^{(\mathrm{ex})}(\bm{\Gamma})=0$.  
Here,
\begin{eqnarray}
\Omega_{\mathrm{eq}}(\bm{\Gamma}) 
&=&
\beta_{\mathrm{eq}}
\dot{\mathcal{H}}(\bm{\Gamma}) - \Lambda(\bm{\Gamma}) 
\nonumber \\
&=&
-\beta_{\mathrm{eq}}
\left[
\dot\gamma V \sigma_{xy}(\bm{\Gamma}) + 2\mathcal{R}(\bm{\Gamma})
\right] 
-\Lambda(\bm{\Gamma}) 
\end{eqnarray}
is the work function for the equilibrium distribution
$\rho_{\mathrm{eq}}(\bm{\Gamma}) =
e^{-\beta_{\mathrm{eq}}\mathcal{H}(\bm{\Gamma})}/\int d\bm{\Gamma} \,
e^{-\beta_{\mathrm{eq}}\mathcal{H}(\bm{\Gamma})}$ with the temperature
$T_{\mathrm{eq}}=\beta_{\mathrm{eq}}^{-1}$.
To proceed, it is convenient to cast $\Omega_{\mathrm{eq}}(\bm{\Gamma})$
in another form.
Note that the shear-stress tensor $\sigma_{xy}(\bm{\Gamma})$ decomposes
into
\begin{equation}
\sigma_{xy}(\bm{\Gamma}) 
=
\sigma_{xy}^{\rm (el)}(\bm{\Gamma})  
+
\sigma_{xy}^{\rm (vis)}(\bm{\Gamma}) ,
\label{eq:sigma_xy_decomp}
\end{equation}
where
\begin{eqnarray}
\sigma_{xy}^{\rm (el)}(\bm{\Gamma})  
&\equiv&
\frac{1}{V}
\sum_{i=1}^{N}
\left[ \, \frac{p_{i,x} p_{i,y} }{ m} + 
y_i F_{i,x}^{{\rm (el)}} \, 
\right],
\label{eq:sigma_xy_el}
\\
\sigma_{xy}^{\rm (vis)}(\bm{\Gamma})  
&\equiv&
\frac{1}{V}
\sum_{i=1}^{N}
y_i F_{i,x}^{{\rm (vis)}}.
\label{eq:sigma_xy_vis}
\end{eqnarray}
From $\bv{F}_{i}^{\rm (vis)} = \sum_{j}' \bv{F}_{ij}^{\rm (vis)} =
- \sum_{j}' \bv{F}_{ji}^{\rm (vis)}$, $\sigma_{xy}^{\rm
(vis)}(\bm{\Gamma})$ can be rewritten as
\begin{eqnarray}
&&
\hspace{-2em}
\sigma_{xy}^{\rm (vis)}(\bm{\Gamma})  
=
\frac{1}{2V} 
\sum_{i,j}{}' y_{ij} F_{ij,x}^{{\rm (vis)}}
\nonumber \\
&&
=
- \frac{\zeta}{2V} 
\sum_{i,j}{}' 
( \bv{v}_{ij} \cdot \hat{\bv{r}}_{ij} ) \, \hat{x}_{ij} \hat{y}_{ij}
r_{ij} 
\Theta(d-r_{ij}).
\label{eq:sigma-decom-03}
\end{eqnarray}
From $\bv{v}_{ij} = \bv{p}_{ij}/m + \bm{\dot\gamma} \cdot \bv{r}_{ij}$
with $\dot\gamma_{\mu \nu} = \dot{\gamma} \delta_{\mu x} \delta_{\nu
y}$, $\sigma_{xy}^{\rm (vis)}(\bm{\Gamma})$ can further be decomposed
into
\begin{equation}
\sigma_{xy}^{\rm (vis)}(\bm{\Gamma})  
=
\sigma_{xy}^{\rm (vis) \, (1)}(\bm{\Gamma})  
+
\sigma_{xy}^{\rm (vis) \, (2)}(\bm{\Gamma}) ,
\label{eq:sigma_xy2}
\end{equation}
where
\begin{eqnarray}
\hspace{-3em}
\sigma_{xy}^{\rm (vis) \, (1)} (\bm{\Gamma}) 
&\equiv&
- \frac{\zeta}{2V}
\sum_{i,j}{'} 
\left( \frac{\bv{p}_{ij}}{m} \cdot \hat{\bv{r}}_{ij} \right)
\hat{x}_{ij} \hat{y}_{ij} r_{ij} 
\Theta(d-r_{ij}),
\label{eq:sigma_xy_vis1}
\hspace{2em}
%\nonumber \\ \\
\\
\sigma_{xy}^{\rm (vis) \, (2)} (\bm{\Gamma}) 
&\equiv&
- \frac{\dot{\gamma} \zeta}{2V}
\sum_{i,j}{'} 
\hat{x}_{ij}^{2} \hat{y}_{ij}^{2} r_{ij}^{2}
\Theta(d-r_{ij}).
\label{eq:sigma_xy_vis2}
\end{eqnarray}
Similarly, ${\cal R}(\bm{\Gamma})$ defined in Eq.~(\ref{app:eq:R}) can
be decomposed into
\begin{equation}
{\cal R}(\bm{\Gamma}) = 
{\cal R}^{(1)}(\bm{\Gamma}) + {\cal R}^{(2)}(\bm{\Gamma}) + {\cal R}^{(3)}(\bm{\Gamma}),
\end{equation}
with
\begin{eqnarray}
{\cal R}^{(1)} (\bm{\Gamma}) 
&\equiv&
\frac{\zeta}{4}
\sum_{i,j}{}' 
\left( \frac{\bv{p}_{ij}}{m}  \cdot \hat{\bv{r}}_{ij} \right)^{2} 
\Theta(d-r_{ij}),
\label{eq:R1-def}
\\
{\cal R}^{(2)} (\bm{\Gamma}) 
&\equiv&
\frac{\dot{\gamma}\zeta}{2}
\sum_{i,j}{}' 
\left( \frac{\bv{p}_{ij}}{m} \cdot \hat{\bv{r}}_{ij} \right)
\hat{x}_{ij} \hat{y}_{ij} r_{ij} 
\Theta(d-r_{ij}),
\hspace{1.5em}
\\
{\cal R}^{(3)} (\bm{\Gamma}) 
&\equiv&
\frac{\dot{\gamma}^{2}\zeta}{4}
\sum_{i,j}{}' 
\hat{x}_{ij}^{2} \hat{y}_{ij}^{2} r_{ij}^{2}
\Theta(d-r_{ij}).
\label{eq:R3-def}
\end{eqnarray}
One easily recognizes from these expressions the following equalities,
\begin{eqnarray}
{\cal R}^{(2)} (\bm{\Gamma}) 
&=&
- \dot{\gamma} V \sigma_{xy}^{\rm (vis) \, (1)}(\bm{\Gamma}),
\\
{\cal R}^{(3)} (\bm{\Gamma}) 
&=&
- \frac{\dot{\gamma}}{2} V \sigma_{xy}^{\rm (vis) \, (2)}(\bm{\Gamma}).
\end{eqnarray}
Using these results, $\Omega_{\mathrm{eq}}(\bv{\Gamma})$ can be
expressed as
\begin{equation}
\Omega_{\mathrm{eq}}(\bm{\Gamma})
\hspace{-0.2em}
=
\hspace{-0.2em}
-\beta_{\mathrm{eq}} \hspace{-0.2em}\left[
\dot\gamma V \sigma_{xy}^{(\mathrm{el})}(\bm{\Gamma})
-
\dot\gamma V \sigma_{xy}^{(\mathrm{vis})(1)}(\bm{\Gamma})
+
2\Delta \mathcal{R}_{\mathrm{eq}}^{(1)}(\bm{\Gamma})
\right],
\end{equation}
where we have defined
\begin{eqnarray}
%\sigma_{xy}^{(\mathrm{el})}(\bm{\Gamma}) 
%&\equiv&
%\frac{1}{V}
%\sum_{i=1}^{N}
%\left[
%\frac{p_i^x p_i^y}{m}
%+
%y_i F_{i}^{(\mathrm{el})x}
%\right],
%\\
%%
%\sigma_{xy}^{(\mathrm{vis})(1)}(\bm{\Gamma})
%&\equiv&
%-\frac{\zeta}{2V}
%\sum_{i,j}{}^{'}
%\hspace{-0.3em}
%\left( \frac{\bm{p}_{ij}}{m}\cdot\hat{\bm{r}}_{ij}\right)
%\Theta(d-r_{ij})
%\hat{x}_{ij}\hat{y}_{ij}r_{ij},
%\\
%%
\Delta \mathcal{R}_{\mathrm{eq}}^{(1)}(\bm{\Gamma})
&\equiv&
\mathcal{R}^{(1)}(\bm{\Gamma})
+
\frac{T_{\mathrm{eq}}}{2}
\Lambda(\bm{\Gamma}).
%\\
%%
%\mathcal{R}^{(1)}(\bm{\Gamma})
%&\equiv&
%\frac{\zeta}{4}
%\sum_{i,j}{}^{'}
%\hspace{-0.3em}
%\left( \frac{\bm{p}_{ij}}{m}\cdot\hat{\bm{r}}_{ij}\right)^2
%\Theta(d-r_{ij}).
\end{eqnarray}

We attempt to obtain an approximate expression of
Eq.~(\ref{app:eq:rho_SS_ex}) in the form of the canonical distribution
and its correction.
The expression Eq.~(\ref{app:eq:rho_SS_ex}) is a product of the
canonical term and the exponential of the time integral of the work
function.
%
%Hence, we approximate the exponential factor as
From Appendix~\ref{app:tauH}, the exponential factor can be rewritten
as
\begin{eqnarray}
\exp
\left[
\int_{-\infty}^{0} d\tau \,
\Omega_{\mathrm{eq}}(\bm{\Gamma}(-\tau))
\right]
\approx
e^{\tau_{\mathrm{rel}}\Omega_{\mathrm{SS}}(\bm{\Gamma})},
\label{app:eq:relaxation_time_approx}
\end{eqnarray}
where $\tau_{\mathrm{rel}}$ is the relaxation time,
Eq.~(\ref{eq:tau_rel_hard-core}).
%
%Note that we have assumed that the time-reversal symmetry,
%$\Omega_{\mathrm{eq}}(\bm{\Gamma}(-\tau)) \approx
%\Omega_{\mathrm{eq}}(\bm{\Gamma}(\tau))$, holds for the case of weak
%dissipation.
%
%
In Eq.~(\ref{app:eq:relaxation_time_approx}), we have defined
\begin{equation}
\Omega_{\mathrm{SS}}(\bm{\Gamma}) 
\equiv
-\beta_{\mathrm{SS}}
\hspace{-0.2em}
\left[
\dot\gamma V \sigma_{xy}^{(\mathrm{el})}(\bm{\Gamma})
-
\dot\gamma V \sigma_{xy}^{(\mathrm{vis})(1)}(\bm{\Gamma})
+
2\Delta \mathcal{R}_{\mathrm{SS}}^{(1)}(\bm{\Gamma})
\right],
\label{app:eq:Omega_SS}
\end{equation}
which is equivalent to $\Omega_{\mathrm{eq}}(\bm{\Gamma})$ except that
the inverse equilibrium temperature $\beta_{\mathrm{eq}}$ is replaced by
its steady-state value $\beta_{\mathrm{SS}}$ via Eq.~(\ref{eq:Hdot}).
Accordingly, we replace $\beta_{\mathrm{eq}}$ in the canonical term of
Eq.~(\ref{app:eq:rho_SS_ex}) by $\beta_{\mathrm{SS}}$.
This gurantees the independence of the steady-state average
\begin{eqnarray}
\langle \cdots \rangle_{\mathrm{SS}} 
\equiv
\int d\bm{\Gamma}\,
\rho_{\mathrm{SS}}(\bm{\Gamma})
\cdots
\end{eqnarray}
on $\beta_{\mathrm{eq}}$.
As a result, we obtain
\begin{eqnarray}
\rho_{\mathrm{SS}}(\bm{\Gamma})
=
\frac{e^{-I_{\mathrm{SS}}(\bm{\Gamma})}}
{\int d\bm{\Gamma} e^{-I_{\mathrm{SS}}(\bm{\Gamma})}},
\label{app:eq:rho_SS}
\end{eqnarray}
where
\begin{eqnarray}
I_{\mathrm{SS}}(\bm{\Gamma})
=
\beta_{\mathrm{SS}} 
\mathcal{H}(\bm{\Gamma}) 
-\tau_{\mathrm{rel}} \Omega_{\mathrm{SS}}(\bm{\Gamma}).
\end{eqnarray}
The steady-state temperature $T_{\mathrm{SS}}=\beta_{\mathrm{SS}}^{-1}$
is determined by the energy balance condition, which is given from
Eq.~(\ref{app:eq:ss}) as
\begin{equation}
\left\langle
\dot{\mathcal{H}}(\bm{\Gamma})
\right\rangle_{\mathrm{SS}}
=
-\dot{\gamma} V 
\left\langle
\sigma_{xy}(\bm{\Gamma})
\right\rangle_{\mathrm{SS}}
-2
\left\langle
\mathcal{R}(\bm{\Gamma})
\right\rangle_{\mathrm{SS}}
= 0.
\label{eq:ss}
\end{equation}

Next, we consider the scaling with respect to $\epsilon$ and evaluate
the order of magnitude of the terms in
$\rho_{\mathrm{SS}}(\bm{\Gamma})$. 
The three terms in Eq.~(\ref{app:eq:Omega_SS}) exhibit the following
scaling properties,
\begin{eqnarray}
\sigma_{xy}^{\rm (el)*}(\bm{\Gamma}) 
&=&
\epsilon^{1/2}
\tilde{\sigma}_{xy}^{\rm (el)}(\bm{\Gamma}), 
\\
\sigma_{xy}^{\rm (vis)(1)*}(\bm{\Gamma}) 
&=&
\epsilon^{3/2}
\tilde{\sigma}_{xy}^{\rm (vis)(1)}(\bm{\Gamma}), 
\\
\mathcal{R}^{(1)*}(\bm{\Gamma})
&=&
\epsilon\,
\tilde{\mathcal{R}}^{(1)}(\bm{\Gamma}),
\\
\Lambda^{*}(\bm{\Gamma})
&=&
\epsilon\,
\tilde{\Lambda}(\bm{\Gamma}),
\end{eqnarray}
where the scaled quantites are given by
\begin{eqnarray}
\tilde{\sigma}_{xy}^{(\mathrm{el})}(\bm{\Gamma})
&=&
\frac{1}{V^*} 
\sum_{i=1}^{N}
\left[
\tilde{p}_{i,x} \tilde{p}_{i,y}
+
\tilde{y}_i \tilde{F}_{i,x}^{(\mathrm{el})}
\right], 
\label{eq:sxy_scaled}
\\
\tilde{\sigma}_{xy}^{\rm (vis)(1)}(\bm{\Gamma}) 
\hspace{-0.2em}
&=&
\hspace{-0.2em}
-\frac{1}{2V^*}
\sum_{i,j}{}^{'} 
\left( \bm{p}_{ij}^{*} \cdot \hat{\bv{r}}_{ij} \right)
\tilde{\hat{x}}_{ij} \tilde{\hat{y}}_{ij} r_{ij}^{*} 
\Theta(1-r_{ij}^{*}),
\hspace{2.5em}
\\
\tilde{\mathcal{R}}^{(1)}(\bm{\Gamma})
&=&
\frac{1}{4}
\sum_{i,j}{}^{'} 
\left( \bm{p}_{ij}^{*} \cdot \hat{\bv{r}}_{ij} \right)^{2} 
\Theta(1-r_{ij}^{*}),
\\
\tilde{\Lambda}(\bm{\Gamma})
&=&
-\sum_{i,j}{}^{'}
\Theta(1-r_{ij}^{*}).
\end{eqnarray}
%
%%%
%
From this, the order of magnitude of the three terms in
$\Omega_{\mathrm{SS}}(\bm{\Gamma})$ can be evaluated as
\begin{eqnarray}
\dot\gamma^* V^* 
\sigma_{xy}^{(\mathrm{el})*}(\bm{\Gamma})
&=&
\epsilon\,
\tilde{\dot\gamma} V^*
\tilde{\sigma}_{xy}^{(\mathrm{el})}(\bm{\Gamma}),
\\
\dot\gamma^* V^*
\sigma_{xy}^{(\mathrm{vis})(1)*}(\bm{\Gamma})
&=&
\epsilon^2\, \tilde{\dot\gamma} V^*
\tilde{\sigma}_{xy}^{(\mathrm{vis})(1)}(\bm{\Gamma}),
\\
\mathcal{R}^{(1)*}(\bm{\Gamma})
&=&
\epsilon\,
\tilde{\mathcal{R}}^{(1)}(\bm{\Gamma}).
\end{eqnarray}
We retain $\sigma_{xy}^{(\mathrm{el})*}(\bm{\Gamma})$,
$\mathcal{R}^{(1)*}(\bm{\Gamma})$ and neglect the higher-order term
$\sigma_{xy}^{(\mathrm{vis})(1)*}(\bm{\Gamma})$, i.e.
\begin{eqnarray}
\Omega_{\mathrm{SS}}^{*}(\bm{\Gamma}) 
\approx
-\epsilon\,\beta_{\mathrm{SS}}^{*}
\hspace{-0.2em}
\left[
\tilde{\dot\gamma} V^{*} \tilde{\sigma}_{xy}^{(\mathrm{el})}(\bm{\Gamma})
+
2\Delta \tilde{\mathcal{R}}_{\mathrm{SS}}^{(1)}(\bm{\Gamma})
\right]. 
\end{eqnarray}
As for the relaxation time, we introduce a scaled relaxation time
$\tilde{\tau}_{\mathrm{rel}}$ with
$
\tau_{\mathrm{rel}}^{*}
=
\epsilon^{-1}
\tilde{\tau}_{\mathrm{rel}},
$
i.e.
\begin{eqnarray}
\tilde{\tau}_{\mathrm{rel}}
=
\left[
\frac{2\sqrt{\pi}}{3} \omega_{E}^{*}(T^{*}_{\mathrm{SS}})
\right]^{-1}.
\label{eq:tau_rel_scaled}
\end{eqnarray}
Note that $\tilde{\tau}_{\mathrm{rel}}$ is related to
$\tilde{\dot\gamma}$ via $T_{\mathrm{SS}}^{*}$, whose explicit
expression will be given later in Eq.~(\ref{app:eq:Tss}).
Then, $I_{\mathrm{SS}}(\bm{\Gamma})$ is approximated as
\begin{eqnarray}
I_{\mathrm{SS}}^{*}(\bm{\Gamma}) 
&\approx&
\beta_{\mathrm{SS}}^{*}
\mathcal{H}^{*}(\bm{\Gamma})
-
\tilde{\tau}_{\mathrm{rel}}
\tilde{\Omega}_{\mathrm{SS}}(\bm{\Gamma}),
\end{eqnarray}
where 
$
\tilde{\Omega}_{\mathrm{SS}}(\bm{\Gamma})
=
-\beta_{\mathrm{SS}}^{*}
\left[
\tilde{\dot\gamma} V^*
\tilde{\sigma}_{xy}^{(\mathrm{el})}(\bm{\Gamma})
+
2\Delta\tilde{\mathcal{R}}^{(1)}_{\mathrm{SS}}(\bm{\Gamma})
\right].
$
%
%
%Although $I^{(\mathrm{nc})}(\bm{\Gamma})$ might be of the same order as
%the canonical term,
%$\beta_{\mathrm{SS}}^{*}\mathcal{H}^{*}(\bm{\Gamma})$, 
%
We treat $\tilde{\tau}_{\mathrm{rel}}
\tilde{\Omega}_{\mathrm{SS}}(\bm{\Gamma})$ as a correction, and expand
$\rho_{\mathrm{SS}}(\bm{\Gamma})$ as
\begin{eqnarray}
\rho_{\mathrm{SS}}(\bm{\Gamma}) 
\approx
\frac{
e^{- \beta_{\mathrm{SS}}^{*} \mathcal{H}^{*}(\bm{\Gamma})}
\left[
1
+
\tilde{\tau}_{\mathrm{rel}}
\tilde{\Omega}_{\mathrm{SS}}(\bm{\Gamma})
\right]
}
{\mathcal{Z}}
\label{app:eq:rho_SS}
\end{eqnarray}
with
$
\mathcal{Z}
\approx
\int d\bm{\Gamma} \,
e^{- \beta_{\mathrm{SS}}^{*} \mathcal{H}^{*}(\bm{\Gamma})}
\left[
1
+
\tilde{\tau}_{\mathrm{rel}}
\tilde{\Omega}_{\mathrm{SS}}(\bm{\Gamma})
\right].
$
%
%where we neglect terms of order $\mathcal{O}(\epsilon)$.
%
An approximate expression for the ensemble average of an observable
$A(\bm{\Gamma})$ with the weight $\rho_{\mathrm{SS}}(\bm{\Gamma})$ can
be obtained from Eq.~(\ref{app:eq:rho_SS}) as
\begin{eqnarray}
\left\langle
A(\bm{\Gamma})
\right\rangle_{\mathrm{SS}}
&\approx&
\left\langle
A(\bm{\Gamma})
\right\rangle_{\mathrm{eq}}
+
\tilde{\tau}_{\mathrm{rel}}
\left\langle
A(\bm{\Gamma})
\tilde{\Omega}_{\mathrm{SS}}(\bm{\Gamma})
\right\rangle_{\mathrm{eq}},
\hspace{0.5em}
\label{app:eq:A_ave}
\end{eqnarray}
where
$
\left\langle
\cdots
\right\rangle_{\mathrm{eq}}
=
\int d\bm{\Gamma} \,
e^{- \beta_{\mathrm{SS}}^{*} \mathcal{H}^{*}(\bm{\Gamma})}
\cdots
$
is the average with respect to the canonical distribution at the
temperature $T_{\mathrm{SS}}$.

\section{Shear viscosity and temperature}

The steady-state average of the shear stress and the energy dissipation
rate is obtained by the formula Eq.~(\ref{app:eq:A_ave}), from which we
can derive an explicit expression for $T_{\mathrm{SS}}$.
From the scaling arguments, the leading contribution to the shear stress
comes from the elastic component
$\sigma_{xy}^{(\mathrm{el})}(\bm{\Gamma})$.
Hence, $\langle \sigma_{xy}(\bm{\Gamma})\rangle_{\mathrm{SS}}$ is
approximately given by
\begin{eqnarray}
\left\langle
\tilde{\sigma}_{xy}(\bm{\Gamma})
\right\rangle_{\mathrm{SS}}
\approx
-
\tilde{\tau}_{\mathrm{rel}} \tilde{\dot\gamma}
\beta_{\mathrm{SS}}^{*} V^{*}
\left\langle
\tilde{\sigma}_{xy}^{(\mathrm{el})}(\bm{\Gamma})
\tilde{\sigma}_{xy}^{(\mathrm{el})}(\bm{\Gamma})
\right\rangle_{\mathrm{eq}}.
\hspace{1em}
\end{eqnarray}
This corresponds to the evaluation of the Green-Kubo formula by the
relaxation time approximation in the correlation function.
Similarly, the leading contribution to the energy dissipation rate comes
from $\mathcal{R}^{(1)}(\bm{\Gamma})$ and hence we obtain
\begin{equation}
\left\langle
\tilde{\mathcal{R}}(\bm{\Gamma})
\right\rangle_{\mathrm{SS}}
\approx
\left\langle
\tilde{\mathcal{R}}^{(1)}(\bm{\Gamma})
\right\rangle_{\mathrm{eq}}
-
2\tilde{\tau}_{\mathrm{rel}} \beta_{\mathrm{SS}}^{*} 
\left\langle
\tilde{\mathcal{R}}^{(1)}(\bm{\Gamma})
\Delta\tilde{\mathcal{R}}_{\mathrm{SS}}^{(1)}(\bm{\Gamma})
\right\rangle_{\mathrm{eq}}.
\end{equation}
In these expressions, the anisotropic terms $\langle
\tilde{\sigma}_{xy}^{(\mathrm{el})}(\bm{\Gamma})\rangle_{\mathrm{eq}}$
and $\langle \tilde{\mathcal{R}}^{(1)}(\bm{\Gamma})
\tilde{\sigma}_{xy}^{(\mathrm{el})}(\bm{\Gamma}) \rangle_{\mathrm{eq}} $
are identically zero.
We obtain the following results in the hard-core limit from
straightforward calculations,
\begin{eqnarray}
&&
\left\langle
\tilde{\mathcal{R}}^{(1)}(\bm{\Gamma})
\right\rangle_{\mathrm{eq}}
=
\frac{\sqrt{\pi}}{2} N T_{\mathrm{SS}}^{*}
\omega_{E}^{*}(T_{\mathrm{SS}}^{*}),
\label{eq:R1_eq}
\\
&&
\left\langle
\tilde{\mathcal{R}}^{(1)}(\bm{\Gamma})
\Delta\tilde{\mathcal{R}}_{\mathrm{SS}}^{(1)}(\bm{\Gamma})
\right\rangle_{\mathrm{eq}}
=
\frac{1}{8} N
T_{\mathrm{SS}}^{*2}
\frac{\omega_{E}(T_{\mathrm{SS}}^{*})^{*2}}
{n^{*}g_{0}(\varphi)}
\nonumber \\
&&
\hspace{1em}
\times
\left[ \mathscr{R}_{2}\, + \mathscr{R}_{3}\, n^{*} g_{0}(\varphi)
\right],
\label{eq:R1DR1_eq}
\\
&&
\left\langle
\tilde{\sigma}_{xy}^{(\mathrm{el})}(\bm{\Gamma})
\tilde{\sigma}_{xy}^{(\mathrm{el})}(\bm{\Gamma})
\right\rangle_{\mathrm{eq}}
=
\frac{1}{V^*} n^* T_{\mathrm{SS}}^{*2}
\nonumber \\
&&
\hspace{1em}
\times
\left[ 1 + \mathscr{S}_{2}\, n^{*} g_{0}(\varphi) + \mathscr{S}_{3}\, n^{*2} g_{0}(\varphi)^2 +
\mathscr{S}_{4}\, n^{*3} g_{0}(\varphi)^3 \right],
\hspace{2.5em}
\label{eq:sxysxy_eq}
\end{eqnarray}
where $\mathscr{R}_{2}=1$, $\mathscr{R}_{3}=3\pi/4$,
$\mathscr{S}_{2}=2\pi/15$, $\mathscr{S}_{3}=-\pi^2/20$, and
$\mathscr{S}_{4}=3\pi^3/160$.
Explicit calculations of the equilibrium correlations,
Eqs.~(\ref{eq:R1_eq})--(\ref{eq:sxysxy_eq}), are shown in
Sec.~\ref{app:EqCorr} in detail.
To be specific, Eq.~(\ref{eq:R1_eq}) is derived from
Eq.~(\ref{eq:R1_eq_derive}), Eq.~(\ref{eq:R1DR1_eq}) from
Eqs.~(\ref{eq:R1Dr1_eq_derive1}) and (\ref{eq:R1Dr1_eq_derive2}), and
Eq.~(\ref{eq:sxysxy_eq}) from (\ref{eq:sxysxy_eq_derive1}),
(\ref{eq:sxysxy_eq_derive2}), and (\ref{eq:sxysxy_4body_5}).
Here, the terms with coefficients $\mathscr{R}_{i}$ and
$\mathscr{S}_{i}$ are contributions of the $i$-body correlations, and
the first term of Eq.~(\ref{eq:sxysxy_eq}) is the contribution of the
kinetic stress.
From the energy balance Eq.~(\ref{eq:ss}) of order~$\epsilon$, i.e.
$
-\tilde{\dot\gamma} V^*
\left\langle
\tilde{\sigma}_{xy}(\bm{\Gamma})
\right\rangle_{\mathrm{SS}}
-2
\left\langle
\tilde{\mathcal{R}}(\bm{\Gamma})
\right\rangle_{\mathrm{SS}}
=
0, 
$
we obtain the steady-state temperature $T_{\mathrm{SS}}$ as
\begin{eqnarray}
T_{\mathrm{SS}}^{*}
=
\frac{3\tilde{\dot\gamma}^2}{32\pi}
\frac{S}{R},
\label{app:eq:Tss}
\end{eqnarray}
where $S$ and $R$ are given by
\begin{eqnarray}
S
&=&
1 + \mathscr{S}_{2}\, n^{*} g_{0}(\varphi) + \mathscr{S}_{3}\, n^{*2} g_{0}(\varphi)^2 +
\mathscr{S}_{4}\, n^{*3} g_{0}(\varphi)^3,
\hspace{2em}
\\
R
&=&
n^{*}g_{0}(\varphi) \left[\mathscr{R}_{2}' \, 
+ \mathscr{R}_{3}' \, n^{*}g_{0}(\varphi)
\right],
\end{eqnarray}
with $\mathscr{R}_{2}' = -3/4$, $\mathscr{R}_{3}' = 7\pi/16$.
We further obtain the expression for the shear stress,
\begin{equation}
\left\langle
\tilde{\sigma}_{xy}(\bm{\Gamma})
\right\rangle_{\mathrm{SS}}
=
-
\frac{3}{8\pi}
\tilde{\dot\gamma}\,
T_{\mathrm{SS}}^{* 1/2}
\frac{S}{g_{0}(\varphi)}
%\nonumber \\
%%
%&=&
=
-
\frac{3\sqrt{6}}{64\pi^{3/2}}
\tilde{\dot\gamma}^2
\frac{S^{3/2}}{R^{1/2}g_{0}(\varphi)}.
\label{app:eq:sxy_ss}
\end{equation}
In the vicinity of the jamming point $\varphi_{J}$, $S$ and $R$ can be
approximated as $S\approx \mathscr{S}_{4}\, n^{*3}g_{0}(\varphi)^3$ and
$R\approx \mathscr{R}_{3}' \, n^{*2}g_{0}(\varphi)^2$, respectively.
This leads to the following approximate expressions,
\begin{eqnarray}
&&
T_{\mathrm{SS}}^{*} 
\approx
\frac{3\tilde{\dot\gamma}^2}{32\pi}
\frac{\mathscr{S}_{4}}{\mathscr{R}_{3}'}
n^{*} g_{0}(\varphi)
=
\frac{9\pi}{2240}\tilde{\dot\gamma}^2 n^{*} g_{0}(\varphi),
\\
&&
\left\langle
\tilde{\sigma}_{xy}(\bm{\Gamma})
\right\rangle_{\mathrm{SS}}
\approx
-\frac{9\pi^2}{1280}
\tilde{\dot\gamma}\,
T_{\mathrm{SS}}^{*\,1/2}
n^{*3} g_{0}(\varphi)^2
\nonumber \\
&&
\hspace{2em}
=
- \frac{27\pi^{5/2}}{10240\sqrt{35}}
\tilde{\dot\gamma}^2
n^{* 7/2} g_{0}(\varphi)^{5/2}.
\end{eqnarray}
In the course of the derivation, we have utilized
Eq.~(\ref{eq:tau_rel_scaled}).
Note that this expression agrees with the scaling from the dimensional
analysis and obeys the Bagnold scaling, $ \langle
\tilde{\sigma}_{xy}(\bm{\Gamma}) \rangle_{\mathrm{SS}} \propto
\tilde{\dot\gamma}^2$.
From these expressions, we obtain the scalings 
\begin{eqnarray}
&&
T_{\mathrm{SS}}^{*}
\sim
g_{0}(\varphi) 
\sim 
(\varphi_{J}-\varphi)^{-1},
\label{app:eq:Tss_scaling}
\\
&&
\left\langle
\tilde{\sigma}_{xy}(\bm{\Gamma})
\right\rangle_{\mathrm{SS}}
\sim
T_{\mathrm{SS}}^{*\,1/2}g_{0}(\varphi)^2
\sim
(\varphi_{J}-\varphi)^{-5/2}.
\label{eq:sigma_xy_scaling}
\end{eqnarray}

From Eq.~(\ref{app:eq:sxy_ss}), % and (\ref{eq:sigma_xy_approx}), 
we obtain the expression for the shear viscosity $\eta^{*}=-\langle
\tilde{\sigma}_{xy}(\bm{\Gamma})\rangle_{\mathrm{SS}}/\tilde{\dot\gamma}$,
\begin{equation}
\eta^{*}
%&=&
%\frac{3\sqrt{6}}{64\pi^{3/2}} 
%\epsilon^{-1/2}\, \dot\gamma^{*} 
%\frac{S^{3/2}}{R^{1/2}g_{0}(\varphi)}
%\nonumber \\
%%
\approx
\frac{27\pi^{5/2}}{10240\sqrt{35}}
%\epsilon^{-1/2} 
\dot\gamma^{*} n^{*7/2} g_{0}(\varphi)^{5/2}
\sim
(\varphi_{J}-\varphi)^{-5/2},
%\hspace{1.5em}
\end{equation}
or for $\tilde{\eta}' =
-\langle\tilde{\sigma}_{xy}(\bm{\Gamma})\rangle_{\mathrm{SS}}/(\tilde{\dot\gamma}
\sqrt{T_{\mathrm{SS}}^{*}}) \propto -\langle
\tilde{\sigma}_{xy}(\bm{\Gamma})\rangle_{\mathrm{SS}}/\tilde{\dot\gamma}^{2}
$,
\begin{eqnarray}
\tilde{\eta}'
%&=&
%\frac{3}{8\pi} 
%\frac{S^{3/2}}{R^{1/2}g_{0}(\varphi)}
%%\nonumber \\
%%%
%%&\to&
\approx
\frac{9\pi^{2}}{1280}
n^{*3} g_{0}(\varphi)^{2}
\sim
(\varphi_{J}-\varphi)^{-2}.
\hspace{1em}
\end{eqnarray}

\section{Equilibrium correlations}
\label{app:EqCorr}

We derive the equilibrium correlations,
Eqs.~(\ref{eq:R1_eq})--(\ref{eq:sxysxy_eq}).
In the course of the derivation, the hard-core limit is taken with the
aid of Eqs.~(\ref{eq:Fel_hard-core_nondim}) and
(\ref{eq:Theta_hard-core_nondim}).

First, Eq.~(\ref{eq:R1_eq}) is calculated  as
\begin{eqnarray}
&&
\hspace{-1em}
\left\langle
\tilde{\mathcal{R}}^{(1)}(\bm{\Gamma})
\right\rangle_{\mathrm{eq}}
=
\frac{1}{4}
\left\langle
\sum_{i,j}{}^{'}
(\bm{p}_{ij}^{*}\cdot\hat{\bm{r}}_{ij})^2 \Theta (1-r_{ij}^{*})
\right\rangle_{\mathrm{eq}} 
\nonumber \\
&&
=
\frac{1}{2} T_{\mathrm{SS}}^{*}
\left\langle
\sum_{i,j}{}^{'}
\Theta (1-r_{ij}^{*})
\right\rangle_{\mathrm{eq}} 
\nonumber \\
&&
\approx
\frac{1}{2}N n^{*} T_{\mathrm{SS}}^{*}
\int d^3 \bm{r}^{*} \,
g(r^{*}) \Theta (1-r^{*})
\nonumber \\
&&
\approx
\frac{\sqrt{\pi}}{2}
N T_{\mathrm{SS}}^{*}
\omega_{E}^{*}(T_{\mathrm{SS}}^{*}),
\label{eq:R1_eq_derive}
\end{eqnarray}
where we have utilized $ \langle \sum_{i,j}{}^{'}
\delta(\bm{r}^{*}-\bm{r}_{ij}^{*}) \rangle_{\mathrm{eq}} \approx Nn^{*}
g(r^{*}) $ with $g(r)$ the radial distribution function at equilibrium
in the third equality and Eq.~(\ref{eq:Theta_hard-core_nondim}) in the
last equality.

Next, we deal with Eq.~(\ref{eq:R1DR1_eq}),
\begin{eqnarray}
\left\langle
\tilde{\mathcal{R}}^{(1)}
\Delta\tilde{\mathcal{R}}_{\mathrm{SS}}^{(1)}
\right\rangle_{\mathrm{eq}} 
=
\left\langle
\tilde{\mathcal{R}}^{(1)}
\tilde{\mathcal{R}}^{(1)}
\right\rangle_{\mathrm{eq}} 
+
\frac{T_{\mathrm{SS}}^{*}}{2}
\left\langle
\tilde{\mathcal{R}}^{(1)}
\tilde{\Lambda}
\right\rangle_{\mathrm{eq}}. 
\hspace{2em}
\label{app:eq:R1DR1}
\end{eqnarray}
%
%%%
%
\begin{figure}[tbh]
%\begin{center}
\includegraphics[height=6cm]{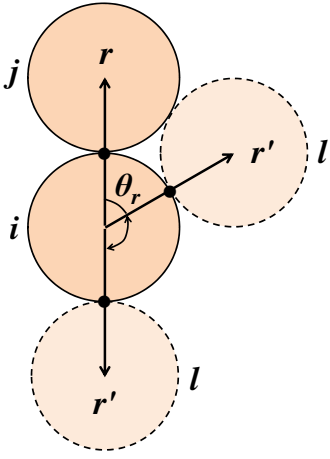} 
%\end{center} 
%\vspace{-2.0em}
\caption {(Color online) Schematic figure for the three-body
 correlation. The angle $\theta_{r}$ between $\hat{\bm{r}}$ and
 $\hat{\bm{r}}'$ is in the range [$\pi/3$, $\pi$].}
\label{Fig:3body}
%\vspace{-1em}
\end{figure}
%
%%%
%
The first term is decomposed as
\begin{eqnarray}
&&
\left\langle
\tilde{\mathcal{R}}^{(1)}
\tilde{\mathcal{R}}^{(1)}
\right\rangle_{\mathrm{eq}} 
=
\frac{T_{\mathrm{SS}}^{*2}}{4}
\left\langle
\sum_{i,j}{}^{'}
\sum_{l,k}{}^{'}
\tilde{\Lambda}(\bm{r}^{*}_{ij})
\tilde{\Lambda}(\bm{r}^{*}_{lk})
\right\rangle_{\mathrm{eq}} 
\nonumber \\
&&
\hspace{1em}
+ 8T_{\mathrm{SS}}^{*2}
\left\langle
\sum_{i,j,l}{}^{''}
\tilde{\mathcal{R}}^{(1)\mu\nu}(\bm{r}_{ij}^{*})
\tilde{\mathcal{R}}^{(1)\mu\nu}(\bm{r}_{il}^{*})
\right\rangle_{\mathrm{eq}} 
\nonumber \\
&&
\hspace{1em}
+ 8T_{\mathrm{SS}}^{*2}
\left\langle
\sum_{i,j}{}^{'}
\tilde{\mathcal{R}}^{(1)\mu\nu}(\bm{r}_{ij}^{*})
\tilde{\mathcal{R}}^{(1)\mu\nu}(\bm{r}_{ij}^{*})
\right\rangle_{\mathrm{eq}},
\end{eqnarray}
where we have defined
$
\tilde{\mathcal{R}}^{(1)\mu\nu}(\bm{r}^{*})
=
\Theta(1-r^{*})
\hat{r}^{\mu}
\hat{r}^{\nu}/4.
$
Here, $\sum_{i,j,l}{}^{''}$ is performed under the condition that any
two pair of particles $(i,j,l)$ is different.
On the other hand, the second term on the right hand side of
Eq.~(\ref{app:eq:R1DR1}) is given by
\begin{eqnarray}
\frac{T_{\mathrm{SS}}^{*}}{2}
\left\langle
\tilde{\mathcal{R}}^{(1)}
\tilde{\Lambda}
\right\rangle_{\mathrm{eq}}
&=&
-\frac{T_{\mathrm{SS}}^{*2}}{4}
\left\langle
\sum_{i,j}{}^{'}
\sum_{l,k}{}^{'}
\tilde{\Lambda}(\bm{r}_{ij}^{*})
\tilde{\Lambda}(\bm{r}_{lk}^{*})
\right\rangle_{\mathrm{eq}}
\hspace{-1em}
.
\hspace{2em}
\end{eqnarray}
Hence, the four-body correlation cancels out and the two- and three-body
correlations remain.

The three-body correlation is calculated as
\begin{eqnarray}
&&
\hspace{-1em}
\left\langle
\sum_{i,j,l}{}^{''}
\tilde{\mathcal{R}}^{(1)\mu\nu}(\bm{r}_{ij}^{*})
\tilde{\mathcal{R}}^{(1)\mu\nu}(\bm{r}_{il}^{*})
\right\rangle_{\mathrm{eq}} 
\nonumber \\
&&
=
\frac{1}{16}
\int d^3 \bm{r}^{*} \int d^3 \bm{r}'{}^{*} 
\left\langle
\sum_{i,j,l}{}^{''}
\delta(\bm{r}^{*}-\bm{r}_{ij}^{*})
\delta(\bm{r}'{}^{*}-\bm{r}_{il}^{*})
\right\rangle_{\mathrm{eq}}
\nonumber \\
&&
\hspace{1em}
\times
\left( \hat{\bm{r}}\cdot\hat{\bm{r}}'\right)^2
\Theta(1-r^{*})
\Theta(1-r'{}^{*})
\nonumber \\
&&
\approx
\frac{N n^{*2}}{16}
\int d^3 \bm{r}^{*} \int d^3 \bm{r}'{}^{*} 
g^{(3)}(\bm{r}^{*},\bm{r}'{}^{*})
\left( \hat{\bm{r}}\cdot\hat{\bm{r}}'\right)^2
\nonumber \\
&&
\hspace{1em}
\times
\Theta(1-r^{*})
\Theta(1-r'{}^{*}),
\end{eqnarray}
where $g^{(3)}(\bm{r},\bm{r}')$ is the triple correlation function at
equilibrium. 
The triple correlation is conventionally approximated by the Kirkwood's
superposition approximation as $g^{(3)}(\bm{r},\bm{r}') \approx
g(r)g(r')g(|\bm{r}-\bm{r}'|)$, where $g(r)$ and $g(r')$ are the radial
correlations and $g(|\bm{r}-\bm{r}'|)$ is the angular
correlation~\cite{HM} (cf. Fig.~\ref{Fig:3body}).
In the present case, we are interested in the case where the spheres
($i$, $j$) and ($i$, $l$) are in contact.
To ensure the connectivity, we insert a step function as
\begin{eqnarray}
g^{(3)}(\bm{r},\bm{r}')
&\approx&
g(r)\Theta(d-r)g(r')\Theta(d-r')g(|\bm{r}-\bm{r}'|), 
\hspace{2em}
\label{eq:g3_approx}
\end{eqnarray}
which leads to 
\begin{eqnarray}
&&
\hspace{-1.0em}
\left\langle
\sum_{i,j,l}{}^{''}
\tilde{\mathcal{R}}^{(1)\mu\nu}(\bm{r}_{ij}^{*})
\tilde{\mathcal{R}}^{(1)\mu\nu}(\bm{r}_{il}^{*})
\right\rangle_{\mathrm{eq}} 
\nonumber \\
&&
\approx
\frac{N n^{*2}}{16}
\int_{0}^{\infty} dr^{*} \int_{0}^{\infty} dr'{}^{*} 
g(r^{*}) g(r'{}^{*})
r^{*2} r'{}^{*2}
\nonumber \\
&&
\hspace{1.0em}
\times
\Theta(1-r^{*})^2
\Theta(1-r'{}^{*})^2
\nonumber \\
&&
\hspace{1.0em}
\times
\int d\mathcal{S} \int d\mathcal{S}'
\left( \hat{\bm{r}}\cdot\hat{\bm{r}}'\right)^2
g(|\bm{r}^{*}-\bm{r}'{}^{*}|).
\label{eq:R1R1}
\end{eqnarray}
Here, $\int d\mathcal{S}$ and $\int d\mathcal{S}'$ represent the angular
integrations with respect to the solid angles of $\hat{\bm{r}}$ and
$\hat{\bm{r}}'$, respectively.
It should be noted that there is a subtlety in the integral of the
angular correlation, $g(|\bm{r}^{*}-\bm{r}'{}^{*}|)$, since spheres $j$
and $l$ are in contact when $\theta_{r}=\pi/3$, where $\cos\theta_{r}
\equiv \hat{\bm{r}}\cdot\hat{\bm{r}}'$.
The radial distribution function $g(r)=1+h(r)$, where $h(r)$ is the
total correlation function, consists of a $\delta$-function (contact)
contribution at $r\approx d$ and a regular contribution, which is
approximately 1.
Hence, it is reasonable to approximate $h(r)$ by the $\delta$-function
contribution, which is given by
\begin{eqnarray}
h(r)
\approx
\frac{1}{4\varphi \delta}
\left[
\frac{6A}{\left( \frac{r/d-1}{\delta} + C\right)^4}
+
\frac{B}{\left( \frac{r/d-1}{\delta} + C\right)^2}
\right], 
\hspace{1em}
\end{eqnarray}
with numerically fitted coefficients, $A\approx 3.43$, $B\approx 1.45$,
and $C\approx 2.25$~\cite{DTS2005}.
Here, $\delta > 0$ is defined by $\varphi = \varphi_{J}(1-\delta)^3$,
which is approximated as $\delta \approx
(\varphi_{J}-\varphi)/(3\varphi_{J})$ for $\varphi \approx \varphi_{J}$.
The angular integral is evaluated as (cf. Fig.~\ref{Fig:3body})
\begin{eqnarray}
&&
\hspace{-1em}
\int d\mathcal{S} \int d\mathcal{S}'
\left( \hat{\bm{r}}\cdot\hat{\bm{r}}'\right)^2
\left[
1+
h(|\bm{r}^{*}-\bm{r}^{*}{}'|)
\right]
\nonumber \\
&&
=
8\pi^2
\int_{-1}^{1/2} d(\cos\theta_{r}) \cos^2\theta_{r}
\left[ 1 + h\left( \sqrt{2(1-\cos\theta_{r})}\right)\right]
\nonumber \\
&&
=
3\pi^2 + 
\int_{0}^{1} dz \, (z+1)\left[ 1 - \frac{1}{2}(z+1)^2\right]^2
\hat{h}(z),
\label{eq:AngInteg}
\end{eqnarray}
where $\hat{h}(z) = \left[ 6A/(z/\delta+C)^4 + (z/\delta + C)^2\right] /
(4\varphi \delta)$.
Note that $\theta_{r}$ is restricted to $\theta_{r}\in [\pi/3,\pi]$ due
to the constraint that any two pair of particles $(i,j,l)$ is different.
In approaching the jamming point, i.e. $\delta\to 0$, $\hat{h}(z)$
behaves as $\hat{h}(z) \sim \delta^{-1} \left[ (\delta/z)^4 +
(\delta/z)^2 \right]\to 0$ and hence does not contribute to the angular
integral.
%
%%%
%
\begin{figure}[tbh]
%\begin{center}
\includegraphics[height=6cm]{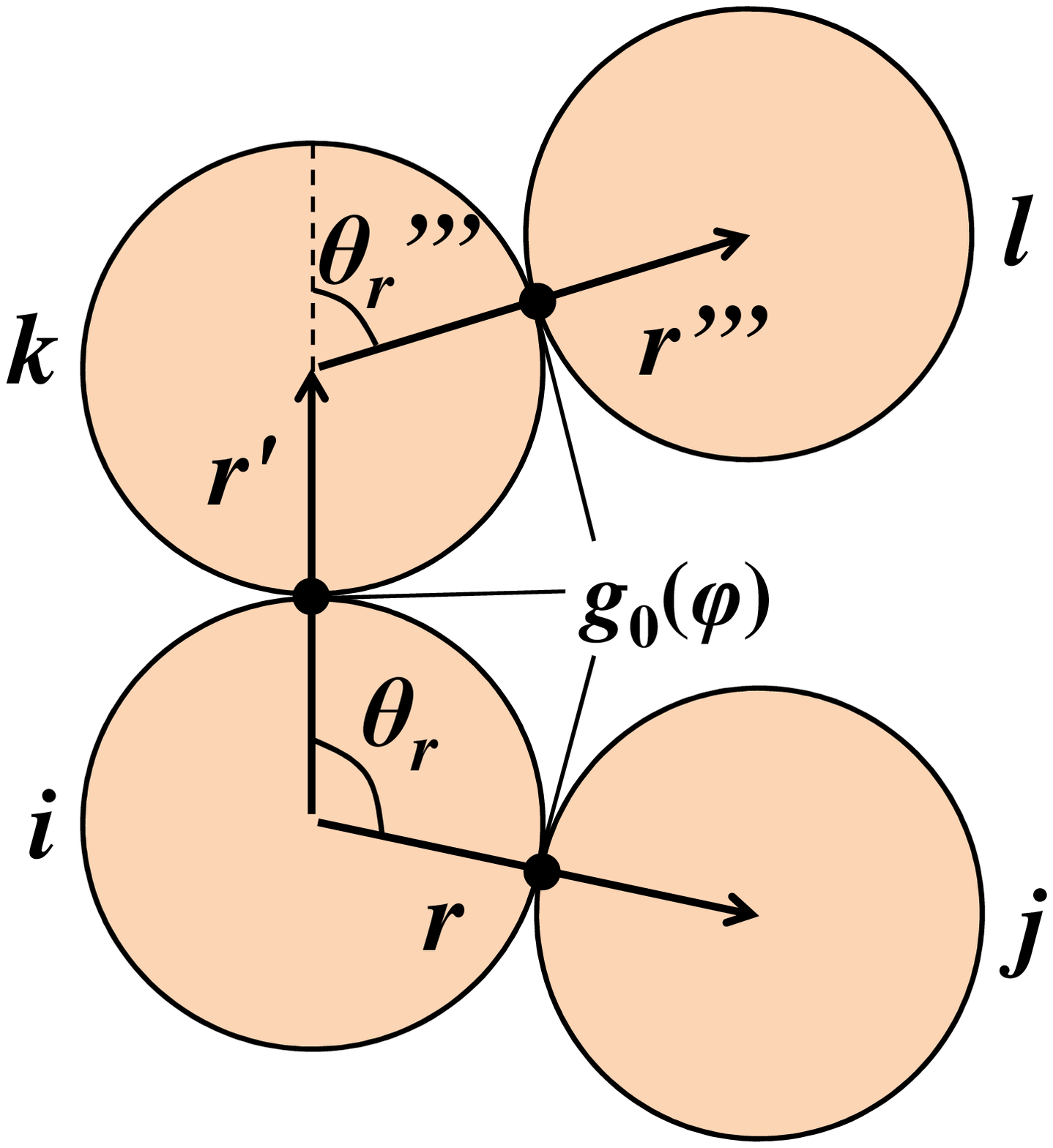} 
%\end{center} 
%\vspace{-2.0em}
\caption {(Color online) Schematic figure for the four-body
 correlation.}
\label{Fig:4body}
%\vspace{-1em}
\end{figure}
%
%%%
%
The radial integrations can be carried out with the aid of
Eq.~(\ref{eq:Theta_hard-core_nondim}), from which we obtain
\begin{equation}
\left\langle
\sum_{i,j,l}{}^{''}
\tilde{\mathcal{R}}^{(1)\mu\nu}(\bm{r}_{ij}^{*})
\tilde{\mathcal{R}}^{(1)\mu\nu}(\bm{r}_{il}^{*})
\right\rangle_{\mathrm{eq}} 
\approx
\frac{3\pi}{256} N \omega_{E}^{*2}(T_{\mathrm{SS}}^{*}). 
\label{eq:R1Dr1_eq_derive1}
\end{equation}
The two-body correlation is calculated, similarly to
Eq.~(\ref{eq:R1_eq_derive}), as
\begin{eqnarray}
&&
\hspace{-2em}
\left\langle
\sum_{i,j}{}^{'}
\tilde{\mathcal{R}}^{(1)\mu\nu}(\bm{r}_{ij}^{*})
\tilde{\mathcal{R}}^{(1)\mu\nu}(\bm{r}_{ij}^{*})
\right\rangle_{\mathrm{eq}} 
\nonumber \\
&=&
\frac{1}{16}
\int d^3 \bm{r}^{*}
\left\langle
\sum_{i,j}{}^{'}
\delta(\bm{r}^{*}-\bm{r}_{ij}^{*})
\right\rangle_{\mathrm{eq}} 
\Theta(1-r^{*})^2
\nonumber \\
&\approx&
\frac{N n^{*}}{16}
g_{0}(\varphi)
\int d^3 \bm{r}^{*}
\Xi(r^{*})^2
=
\frac{1}{64}N
\frac{\omega_{E}^{*2}(T_{\mathrm{SS}}^{*})}{n^{*}g_{0}(\varphi)},
\hspace{2em}
\label{eq:R1Dr1_eq_derive2}
\end{eqnarray}
where Eq.~(\ref{eq:Theta_hard-core_nondim}) is utilized in the last
equality. 
This concludes the derivation of Eq.~(\ref{eq:R1DR1_eq}).

Finally, we deal with Eq.~(\ref{eq:sxysxy_eq}).
From the definition of the elastic stress tensor,
Eq.~(\ref{eq:sxy_scaled}), we obtain
\begin{eqnarray}
&&
\hspace{-1em}
\left\langle
\tilde{\sigma}_{xy}^{(\mathrm{el})}
\tilde{\sigma}_{xy}^{(\mathrm{el})}
\right\rangle_{\mathrm{eq}}  
=
\frac{1}{V^{*2}}
\left\{
N T_{\mathrm{SS}}^{*2}
\right.
\nonumber \\
&&
\left.
+
\frac{1}{4}
\left\langle
\sum_{i,j}{}^{'}
\sum_{l,k}{}^{'}
y_{ij}^{*} y_{lk}^{*}
F_{ij}^{(\mathrm{el})x*}
F_{lk}^{(\mathrm{el})x*}
\right\rangle_{\mathrm{eq}}  
\right\},
\end{eqnarray}
where the first term is the kinetic stress and the second term is the
contact stress.
The contact stress consists of three components, i.e. the two-, three-,
and four-body correlations.

First, the two-body correlation is calculated as
\begin{eqnarray}
&&
\hspace{-1em}
\frac{1}{2}
\left\langle
\sum_{i,j}{}^{'}
y_{ij}^{*2}
F_{ij}^{(\mathrm{el})x*2}
\right\rangle_{\mathrm{eq}}
\approx
\frac{1}{2}N n^{*}
\int d^3 \bm{r}^{*} g(r^{*}) y^{*2}
F_{x}^{(\mathrm{el})*2}
\nonumber \\
&&
=
\frac{1}{2} N n^{*}
\int_{0}^{\infty} dr^{*} g(r^{*})
F^{(\mathrm{el})*2}
r^{*4}
\int d\mathcal{S} \, \hat{x}^{2}\hat{y}^{2} 
\nonumber \\
&&
=
\frac{2\pi}{15} N n^{*}
\int_{0}^{\infty} dr^{*} g(r^{*})
F^{(\mathrm{el})*2} r^{*4},
\label{eq:sxy_2body}
\end{eqnarray}
where $F^{(\mathrm{el})*}(r^{*})= -\Theta(1-r^{*})\,u^{*}{}'(r^{*})$ is
the magnitude of the elastic force with $u^{*}(r^{*})$ the pair
potential, and $\int d\mathcal{S}\cdots$ is the integration with respect
to the solid angle of $\hat{\bm{r}}$.
From Eq.~(\ref{eq:Fel_hard-core_nondim}), we we obtain the following
expression in the hard-core limit,
\begin{eqnarray}
\frac{1}{2}
\left\langle
\sum_{i,j}{}^{'}
y_{ij}^{*2}
F_{ij}^{(\mathrm{el})x*2}
\right\rangle_{\mathrm{eq}}
\approx
\frac{2\pi}{15} N T_{\mathrm{SS}}^{*2}
n^{*} g_{0}(\varphi).
\label{eq:sxysxy_eq_derive1}
\end{eqnarray}
Next, the three-body correlation is calculated as
\begin{eqnarray}
&&
\hspace{-1em}
\frac{1}{2}
\left\langle
\sum_{i,j,k}{}^{''}
y_{ij}^{*} y_{ik}^{*}
F_{ij}^{(\mathrm{el})x*}
F_{ik}^{(\mathrm{el})x*}
\right\rangle_{\mathrm{eq}}
\nonumber \\
&&
\approx
\frac{1}{2} N n^{*2}
\int d^{3}\bm{r}^{*} \int d^3 \bm{r}'{}^{*}
g^{(3)}(\bm{r}^{*},\bm{r}'{}^{*})
r^{*}r'{}^{*}
\nonumber \\
&&
\hspace{1em}
\times
\hat{x}\hat{y} \hat{x}'\hat{y}'
F^{(\mathrm{el})*}(r^{*})
F^{(\mathrm{el})*}(r'{}^{*})
\nonumber \\
&&
\approx
\frac{1}{2} N n^{*2}
\int_{0}^{\infty} dr^{*} \int_{0}^{\infty} dr'{}^{*}
g(r^{*})g(r'{}^{*})
r^{*3}r'{}^{*3}
\nonumber \\
&&
\hspace{1em}
\times
F^{(\mathrm{el})*}(r^{*})
F^{(\mathrm{el})*}(r'{}^{*})
\Theta (1-r^{*})
\Theta (1-r'{}^{*})
\nonumber \\
&&
\hspace{1em}
\times
\int d\mathcal{S} \int d\mathcal{S}'
\hat{x}\hat{y} \hat{x}'\hat{y}' \,
g(|\bm{r}^{*}-\bm{r}^{*}{}'|),
\end{eqnarray}
where Eq.~(\ref{eq:g3_approx}) is applied in the second equality.
Similarly to Eq.~(\ref{eq:R1R1}), the $\delta$-function contribution of
$g(|\bm{r}^{*}-\hat{\bm{r}}^{*}{}'|)$ vanishes in the angular integrals.
Hence, we obtain
\begin{eqnarray}
&&
\hspace{-0.5em}
\int d\mathcal{S} \int d\mathcal{S}' \,
\hat{x}\hat{y} \hat{x}'\hat{y}' \,
g(|\bm{r}^{*}-\bm{r}^{*}{}'|)
\nonumber \\
&&
=
\int_{-1}^{\frac{1}{2}} d(\cos\theta_{r})
\int_{-1}^{1} d(\cos\theta')
\int_{0}^{2\pi} \hspace{-0.5em} d\phi_{r}
\int_{0}^{2\pi} \hspace{-0.5em} d\phi'
\hat{x}\hat{y}\hat{x}'\hat{y}'
\nonumber \\
&&
=
-\frac{\pi^2}{10}.
\label{eq:3body-hc}
\end{eqnarray}
Here, $\cos\theta_{r}\equiv \hat{\bm{r}}\cdot\hat{\bm{r}}'$ and
$\theta'$, $\phi'$ are defined as
\begin{eqnarray}
\hat{x}'
&=&
\sin\theta'\cos\phi',
\label{eq:hatx_prime}
\\
\hat{y}'
&=&
\sin\theta'\sin\phi',
\\
\hat{z}'
&=&
\cos\theta'.
\label{eq:hatz_prime}
\end{eqnarray}
This gives $\hat{x}$ and $\hat{y}$ in terms of $\theta'$, $\phi'$,
$\theta_{r}$, and $\phi_{r}$ as
\begin{eqnarray}
\hat{x}
&=&
\left( 1 - \frac{\hat{x}'{}^{2}}{1+\hat{z}'}\right) \sin\theta_{r}\cos\phi_{r}
-
\frac{\hat{x}'\hat{y}'}{1+\hat{z}'}\sin\theta_{r}\sin\phi_{r}
\nonumber \\
&&
+
\hat{x}'\cos\theta_{r},
\label{eq:xhat}
\\
\hat{y}
&=&
-
\frac{\hat{x}'\hat{y}'}{1+\hat{z}'}\sin\theta_{r}\cos\phi_{r}
+
\left( 1 - \frac{\hat{y}'{}^{2}}{1+\hat{z}'}\right) \sin\theta_{r}\sin\phi_{r}
\nonumber \\
&&
+
\hat{y}'\cos\theta_{r},
\label{eq:yhat}
\end{eqnarray}
%
%%%
%
\begin{figure*}[tbh]
%\begin{center}
\includegraphics[height=6cm]{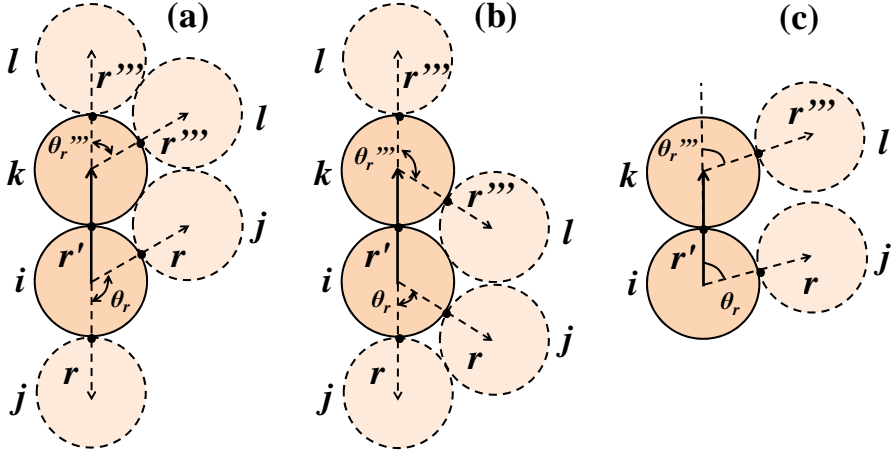} 
%\end{center} 
%\vspace{-2.0em}
\caption {(Color online) Schematic figure for the four-body
 correlation.}
\label{Fig:4body-abc}
%\vspace{-1em}
\end{figure*}
%
%%%
%
\hspace{-1em} where $\phi_{r}$ is the azimuthal angle of
$\hat{\bm{r}}$ in the spherical coordinate where $\hat{\bm{r}}'$ points
in the $z$-direction.
Note that $\theta_{r}$ is restricted to $\theta_{r} \in [\pi/3,\pi]$,
similarly to Eq.~(\ref{eq:AngInteg}).
The radial integrations can be carried out with the aid of
Eq.~(\ref{eq:Fel_hard-core_nondim}), from which we obtain
\begin{equation}
\frac{1}{2}
\left\langle
\sum_{i,j,k}{}^{''}
y_{ij}^{*} y_{ik}^{*}
F_{ij}^{(\mathrm{el})x*}
F_{ik}^{(\mathrm{el})x*}
\right\rangle_{\mathrm{eq}}
\approx
-\frac{\pi^2}{20} N T_{\mathrm{SS}}^{*2} n^{*2} g_{0}(\varphi)^2.
\label{eq:sxysxy_eq_derive2}
\end{equation}
Finally, the four-body correlation is calculated as
\begin{eqnarray}
&&
\hspace{-0.5em}
\frac{1}{4}
\left\langle
\sum_{i,j,l,k}{}^{'''}
y_{ij}^{*} y_{lk}^{*}
F_{ij}^{(\mathrm{el})x*}
F_{lk}^{(\mathrm{el})x*}
\right\rangle_{\mathrm{eq}}
\nonumber \\
&&
=
\frac{1}{4}
\left\langle
\sum_{i,j,l,k}{}^{'''}
y_{ij}^{*} 
(y_{ik}^{*}-y_{il}^{*})
F_{ij}^{(\mathrm{el})x*}
F_{lk}^{(\mathrm{el})x*}
\right\rangle_{\mathrm{eq}}
\nonumber \\
&&
\approx
\frac{1}{2}
N n^{*3}
\int d^3 \bm{r}^{*} \int d^3 \bm{r}'{}^{*} \int d^3 \bm{r}''{}^{*}
g^{(4)}(\bm{r}^{*},\bm{r}'{}^{*},\bm{r}''{}^{*})
\nonumber \\
&&
\times
y^{*}(y'{}^{*}-y''{}^{*})
\hat{x}(\hat{x}'-\hat{x}'')
F^{(\mathrm{el})*}(r^{*})
F^{(\mathrm{el})*}(|\bm{r}'{}^{*}-\bm{r}''{}^{*}|),
\label{eq:sxysxy_4body_1}
\nonumber \\ 
\end{eqnarray}
where $g^{(4)}(\bm{r},\bm{r}',\bm{r}'')$ is the quadruple correlation at
equilibrium. 
Here, $\sum_{i,j,l,k}{}^{'''}$ is performed under the condition that any
two pair of particles $(i,j,l,k)$ is different.
We change the integration variable from $\bm{r}''{}^{*}$ to
$\bm{r}'''{}^{*} \equiv \bm{r}''{}^{*} - \bm{r}'{}^{*}$, which leads to
\begin{eqnarray}
&&
\hspace{-0.5em}
\frac{1}{4}
\left\langle
\sum_{i,j,l,k}{}^{'''}
y_{ij}^{*} y_{lk}^{*}
F_{ij}^{(\mathrm{el})x*}
F_{lk}^{(\mathrm{el})x*}
\right\rangle_{\mathrm{eq}}
\nonumber \\
&&
=
\frac{1}{2}
N n^{*3}
\int d^3 \bm{r}^{*} \int d^3 \bm{r}'{}^{*} \int d^3 \bm{r}'''{}^{*}
g^{(4)}(\bm{r}^{*},\bm{r}'{}^{*},\bm{r}'''{}^{*})
\nonumber \\
&&
\times
y^{*}y'''{}^{*}
\hat{x}\hat{x}'''
F^{(\mathrm{el})*}(r^{*})
F^{(\mathrm{el})*}(r'''{}^{*}).
\label{eq:sxysxy_4body_2}
\end{eqnarray}
Similarly to Eq.~(\ref{eq:g3_approx}), we adopt the following
approximation, 
\begin{eqnarray}
&&
\hspace{-0.5em}
g^{(4)} (\bm{r},\bm{r}',\bm{r}''')
\approx
g(r)\Theta(d-r)
g(r')\Theta(d-r')
g(r''')\Theta(d-r''')
\nonumber \\
&&
\times
g(|\bm{r}-\bm{r}'|) 
g(|\bm{r}'+\bm{r}'''|) 
g(|\bm{r}'+\bm{r}'''-\bm{r}|),
\end{eqnarray}
where the step function is inserted to ensure the connectivity of
the spheres ($i$, $j$), ($k$, $l$), and ($i$, $k$).
Then, Eq.~(\ref{eq:sxysxy_4body_2}) is approximated as
\begin{eqnarray}
&&
\hspace{-0.5em}
\frac{1}{4}
\left\langle
\sum_{i,j,l,k}{}^{'''}
y_{ij}^{*} y_{lk}^{*}
F_{ij}^{(\mathrm{el})x*}
F_{lk}^{(\mathrm{el})x*}
\right\rangle_{\mathrm{eq}}
\nonumber \\
&&
\approx
\frac{1}{2}
N n^{*3}
\int_{0}^{\infty} dr^{*} 
\int_{0}^{\infty} dr'{}^{*} 
\int_{0}^{\infty} dr'''{}^{*}
g(r^{*})g(r'{}^{*})g(r'''{}^{*})
\nonumber \\
&&
\times
r^{*3} r'{}^{*2} r'''{}^{*3}
F^{(\mathrm{el})*}(r^{*})
F^{(\mathrm{el})*}(r'''{}^{*})
\nonumber \\
&&
\times
\Theta (1-r^{*})
\Theta (1-r'{^{*}})
\Theta (1-r'''{}^{*})
\nonumber \\
&&
\times
\int d\mathcal{S} \int d\mathcal{S}' \int d\mathcal{S}'''
\hat{x}\hat{y}
\hat{x}'''\hat{y}'''
\nonumber \\
&&
\times
g(|\bm{r}^{*}-\bm{r}'{}^{*}|) 
g(|\bm{r}'{}^{*}+\bm{r}'''{}^{*}|) 
g(|\bm{r}'{}^{*}+\bm{r}'''^{*}-\bm{r}^{*}|).
\label{eq:sxysxy_4body_3}
\end{eqnarray}
Here, $\int d\mathcal{S}$, $\int d\mathcal{S}'$, $\int d\mathcal{S}'''$
represent the angular integrations with respect to the solid angles of
$\hat{\bm{r}}$, $\hat{\bm{r}}'$, $\hat{\bm{r}}'''$, respectively.
We first consider the radial integration in
Eq.~(\ref{eq:sxysxy_4body_3}). 
From Eq.~(\ref{eq:Fel_hard-core_nondim}), integrations with $r^{*}$ and
$r'''{}^{*}$ give $T_{\mathrm{SS}}^{*2}\, g_{0}(\varphi)^2$.
On the other hand, the integration with $r'{}^{*}$ gives
$g_{0}(\varphi)$.
Hence, the radial integration is given by $T_{\mathrm{SS}}^{*2}
g_{0}(\varphi)^3$. 

Next we consider the angular integration in
Eq.~(\ref{eq:sxysxy_4body_3}).
Similarly to Eq.~(\ref{eq:R1R1}), the $\delta$-function contribution of
the raidal correlations $g(|\bm{r}^{*}-\bm{r}'^{*}|)$,
$g(|\bm{r}'{}^{*}+\bm{r}'''{}^{*}|)$, and
$g(|\bm{r}'{}^{*}+\bm{r}'''{}^{*}-\bm{r}^{*}|)$ vanishes.
Hence, it is given by
\begin{eqnarray}
&&
\hspace{-0.5em}
\int d\mathcal{S} \int d\mathcal{S}' \int d\mathcal{S}'''
\hat{x}\hat{y}
\hat{x}'''\hat{y}'''
\nonumber \\
&&
=
\int_{-1}^{1} d(\cos\theta')
\int d(\cos\theta_{r})
\int d(\cos\theta_{r}''')
\nonumber \\
&&
\times
\int_{0}^{2\pi} d\phi'
\int d\phi_{r}
\int d\phi_{r}''' \,
\hat{x} \hat{y} \hat{x}''' \hat{y}''',
%\hspace{1em}
\label{eq:sxysxy-4body-angular}
\end{eqnarray}
where $\cos\theta_{r} \equiv \hat{\bm{r}}\cdot\hat{\bm{r}}'$,
$\cos\theta_{r}'''\equiv\hat{\bm{r}}'''\cdot\hat{\bm{r}}'$, and
$\theta'$, $\phi'$ are defined by
Eqs.~(\ref{eq:hatx_prime})--(\ref{eq:hatz_prime}).
Then, $\hat{x}$ and $\hat{y}$ are given by Eqs.~(\ref{eq:xhat}) and
(\ref{eq:yhat}), and $\hat{x}'''$ and $\hat{y}'''$ are given in terms of
$\theta_{r}'''$, $\phi_{r}'''$, $\theta'$, and $\phi'$ by the
replacement $\theta_{r} \to \theta_{r}'''$, $\phi_{r} \to \phi_{r}'''$
in Eqs.~(\ref{eq:xhat}) and (\ref{eq:yhat}).
For the case (a), where $\theta_{r} \in [\pi/3, \pi]$ and $\theta_{r}'''
\in [0, \pi/3]$ (cf. Fig.~\ref{Fig:4body-abc}), sphere $j$ and $l$ can
move freely without interference in the $\phi_{r}$ and $\phi_{r}'''$
direction, i.e. $\phi_{r} \in [0, 2\pi]$ and $\phi_{r}''' \in [0,
2\pi]$.
Then, the angular integral is evaluated as
\begin{eqnarray}
\int_{-1}^{\frac{1}{2}} d(\cos\theta_{r})
\int_{\frac{1}{2}}^{1} d(\cos\theta_{r}''') \,
f(\theta_{r},\theta_{r}''')
=
-\frac{3\pi^3}{80},
\end{eqnarray}
where $f(\theta_{r},\theta_{r}''')\equiv
4\pi^3(3\cos^2\theta_{r}-1)(3\cos^2\theta_{r}'''-1)/15$ is the result of
the integration with respect to $\theta'$, $\phi'$, $\phi_{r}$, and
$\phi'''_{r}$.
Similarly, for the case (b), where $\theta_{r} \in [2\pi/3, \pi]$ and $\theta_{r}'''
\in [0, 2\pi/3]$ (cf. Fig.~\ref{Fig:4body-abc}), the angular integral is
evaluated as
\begin{eqnarray}
\int_{-1}^{-\frac{1}{2}} d(\cos\theta_{r})
\int_{-\frac{1}{2}}^{1} d(\cos\theta_{r}''') 
f(\theta_{r},\theta_{r}''')
=
-\frac{3\pi^3}{80}.
\hspace{1em}
\end{eqnarray}
As for the case (c), where $\theta_{r}$ and $\theta_{r}'''$ vary in the
range $\theta_{r} \in [\pi/3, 2\pi/3]$, $\theta_{r}''' \in [\pi/3,
\theta_{r}]$ (cf. Fig.~\ref{Fig:4body-abc}), we obtain
\begin{eqnarray}
\int_{-\frac{1}{2}}^{\frac{1}{2}} d(\cos\theta_{r})
\int_{\cos\theta_{r}}^{\frac{1}{2}} d(\cos\theta_{r}''')
f(\theta_{r},\theta_{r}''')
=
\frac{9\pi^3}{80}. 
\hspace{1em}
\end{eqnarray}
Hence, the angular integration is given by $3\pi^3/80$.
Together with the radial integration, we obtain
\begin{equation}
\frac{1}{4}
\left\langle
\sum_{i,j,l,k}{}^{'''}
y_{ij}^{*} y_{lk}^{*}
F_{ij}^{(\mathrm{el})x*}
F_{lk}^{(\mathrm{el})x*}
\right\rangle_{\mathrm{eq}}
\approx
\frac{3\pi^3}{160}
N T_{\mathrm{SS}}^{*2} n^{*3} g_{0}(\varphi)^3.
\label{eq:sxysxy_4body_5} 
\end{equation}
This concludes the derivation of Eq.~(\ref{eq:sxysxy_eq}).

\section{Hard-core limit}
\label{app:hard-core}

In systems of soft spheres with contact interactions, the Heaviside's
step function, $\Theta(d-r)$, appears in the interparticle forces as in
Eqs.~(\ref{eq:elastic-force}) and (\ref{eq:viscous-force}).
This step function reduces to a delta function, $\delta(d-r)$, in the
hard-core limit.
Here, we derive formulas valid in this limit.

First, we derive a formula for the step function in the elastic force.
It is convenient to introduce a function which is referred to as the
cavity distribution function~\cite{HM},
\begin{eqnarray}
y(r)
&=&
g(r)\, e(r)^{-1},
\end{eqnarray}
where
\begin{eqnarray}
e(r)
&=&
e^{-\beta_{\mathrm{SS}}\,u(r)}
\label{eq:er}
\end{eqnarray}
with $u(r)=(\kappa/2)\,\Theta(d-r)(d-r)^2$ the pair potential of the
elastic force.
It should be noted that $e(r)$ reduces to the step function in the
hard-core limit, $\kappa d^2/T_{\mathrm{SS}}\to\infty$, and hence its
derivative $e'(r)$ reduces to the delta function,
$
e'(r) 
\to
\delta(d-r),
$
or $e'(r^{*})\to\delta(1-r^{*})$ in the non-dimensionalized form.
Let us consider the hard-core limit of 
\begin{eqnarray}
\int_{0}^{\infty} \hspace{-0.5em} dr^{*} 
g(r^{*}) F^{(\mathrm{el})*} r^{*2},
\end{eqnarray}
where $F^{(\mathrm{el})*}(r^{*})= -\Theta(1-r^{*})\,u^{*}{}'(r^{*})$ is
the magnitude of the elastic force with $u^{*}(r^{*})$ the pair
potential, for illustration.
From $u'(r)e(r) = -T_{\mathrm{SS}}\, e'(r)$, we obtain the following
expression
\begin{eqnarray}
&&
%\hspace{-1em} 
\int_{0}^{\infty} \hspace{-0.5em} dr^{*} 
g(r^{*}) F^{(\mathrm{el})*} r^{*2}
=
-\int_{0}^{\infty} \hspace{-0.5em} dr^{*} 
g(r^{*}) \Theta(1-r^{*}) u^{*}{}'(r^{*})
r^{*2}
\nonumber \\
&&
=
-T_{\mathrm{SS}}^{*}
\int_{0}^{\infty} \hspace{-0.5em} dr^{*} 
g(r^{*}) e'(r^{*}) r^{*2}
\to
-T_{\mathrm{SS}}^{*}\,
g_{0}(\varphi).
\end{eqnarray}
This corresponds to the replacement
\begin{eqnarray}
F^{(\mathrm{el})}(r) g(r)
&\to&
-T_{\mathrm{SS}}\, \delta(d-r) g_{0}(\varphi),
\label{eq:Fel_hard-core}
\end{eqnarray}
or, in non-dimensionalized form,
\begin{eqnarray}
F^{(\mathrm{el})*}(r^{*}) g(r^{*})
&\to&
-T_{\mathrm{SS}}^{*}\, \delta(1-r^{*}) g_{0}(\varphi).
\label{eq:Fel_hard-core_nondim}
\end{eqnarray}

Next, we derive a formula for the step function in the viscous force
Eq.~(\ref{eq:viscous-force}), which appears in e.g.
Eqs.~(\ref{eq:Sllod-Lambda}), (\ref{eq:sigma_xy_vis1}), and
(\ref{eq:R1-def}).
In accordance with the elastic force, it is convenient to introduce a
variant of the cavity distribution function,
\begin{eqnarray}
y_{1/2}(r)
&=&
g(r)\, e_{1/2}(r)^{-1},
\label{eq:y12}
\end{eqnarray}
where
\begin{eqnarray}
e_{1/2}(r)
&=&
e^{-\sqrt{2\beta_{\mathrm{SS}}\, u(r)}}.
\end{eqnarray}
Similarly to Eq.~(\ref{eq:er}), $e_{1/2}(r)$ reduces to the step
function $\Theta(d-r)$ in the hard-core limit, and hence its derivative $e_{1/2}'(r)$
converges to the delta function $\delta(d-r)$.
We consider the following quantity, which has dimension of
(time)$^{-1}$, for illustration,
\begin{eqnarray}
&&
\hspace{-1em}
\left\langle
\Lambda(\bm{\Gamma}) 
\right\rangle_{\mathrm{eq}}
=
- \frac{\zeta}{m}
\left\langle
\sum_{i,j}{}^{'}
\Theta(d-r_{ij})
\right\rangle_{\mathrm{eq}}
\nonumber \\
&&
\approx
- \epsilon Nn \sqrt{\frac{\kappa}{m}}
\int d^3 \bm{r}\,
g(r) \Theta(d-r).
\end{eqnarray}
From $e_{1/2}'(r)\,
y_{1/2}(r)=\sqrt{\beta_{\mathrm{SS}}\kappa}\,\Theta(d-r)g(r)$, we obtain
\begin{eqnarray}
&&
\hspace{-1em}
\left\langle
\Lambda(\bm{\Gamma}) 
\right\rangle_{\mathrm{eq}}
=
- \epsilon Nn 
\sqrt{\frac{T_{\mathrm{SS}}}{m}}\sqrt{\frac{\kappa}{T_{\mathrm{SS}}}}
\int d^3 \bm{r}\,
g(r) \Theta(d-r)
\nonumber \\
&&
=
- \epsilon Nn 
\sqrt{\frac{T_{\mathrm{SS}}}{m}}
\int d^3 \bm{r}\,
e_{1/2}'(r) y_{1/2}(r)
\nonumber \\
&&
\to
- 4\pi \epsilon Nn 
\sqrt{\frac{T_{\mathrm{SS}}}{m}}
g_{0}(\varphi) d^2,
\end{eqnarray}
which corresponds to the replacement
\begin{eqnarray}
\sqrt{\frac{\kappa}{m}} 
\Theta(d-r) \, g(r)
\to
\sqrt{\frac{T_{\mathrm{SS}}}{m}}
\delta(d-r) \, g_{0}(\varphi).
\label{eq:Theta_hard-core}
\end{eqnarray}
In the non-dimensionalized form, it is given by
\begin{equation}
\Theta(1-r^{*})\, g(r^{*})
\to
\Xi(r^{*})\, g_{0}(\varphi),
\label{eq:Theta_hard-core_nondim}
\end{equation}
where
\begin{equation}
\Xi(r^{*})
\equiv
\sqrt{T_{\mathrm{SS}}^{*}}\,
\delta(1-r^{*})
=
\frac{\omega_{E}^{*}(T_{\mathrm{SS}}^{*})}{4\sqrt{\pi}\,n^{*}g_{0}(\varphi)}
\delta(1-r^{*}).
\label{eq:Xi}
\end{equation}

\section{Molecular dynamics simulation}

We briefly describe the methods of the molecular dynamics (MD)
simulation we have performed.
The unit of mass, length, and time is chosen as $m$, $d$,
$\sqrt{m/\kappa}$, and we attach $*$ to non-dimenzionalized quantities,
e.g. $t^{*} = t \sqrt{\kappa/m}$.
The parameters of the simulation are the volume fraction $\varphi$, the
shear rate $\dot\gamma^{*}$, the dissipation rate $\epsilon = \zeta
/ \sqrt{\kappa m}$, and the number of spheres $N$.
The conditions are $N=2000$, $\epsilon = 0.018375$,
$\dot\gamma^{*}=10^{-3}, 10^{-4}, 10^{-5}$, while $\varphi$ is varied in
the range 0.50 and 0.66.

The governing equation is the Sllod equation for uniformly sheared
systems, Eq.~(\ref{app:eq:Sllod}).
This equation is integrated numerically by the velocity Verlet algorithm.
The time step of the calculation is chosen as $\Delta t^{*} = 0.01$.
We start with thermalizing the system at an initial temperature
$T_{\mathrm{ini}}$ in the absence of shear and dissipation, i.e. by
setting $\bm{\dot\gamma}=0$ and $\bm{F}^{(\mathrm{vis})}=0$ in
Eq.~(\ref{app:eq:Sllod}).
After thermalization, we switch on the shear and dissipation
simultaneously, and evolve the system until the temperature
$T(t)=\sum_{i=1}^{N} \bm{p}_{i}(t)^2 /(3Nm)$ is relaxed and starts to
fluctuate around a steady value.
We regard this behavior as a signal for reaching the steady state, and
this steady value is identified as the steady-state temperature,
$T_{\mathrm{SS}}$.
Note that $T_{\mathrm{SS}}$ is determined solely by the balance of
energy, i.e. $\dot\gamma$ and $\zeta$, and is independent of the choice
of $T_{\mathrm{ini}}$.
We extract the relaxation time $\tau_{\mathrm{rel}}$ by fitting the
relaxation behavior of $T(t)$ with an exponential function,
$e^{-t/\tau_{\mathrm{rel}}}$.

Then, the ensemble average of the shear stress around the steady state
is measured by calculating
\begin{equation}
\left\langle
\sigma_{xy}^{*}(\bm{\Gamma})
\right\rangle_{\mathrm{SS}}
=
\frac{1}{V^*} \sum_{i=1}^{N}
\left[
p_{i,x}^{*} p_{i,y}^{*}
+
y_{i}^{*} \left( F_{i,x}^{(\mathrm{el})*}+F_{i,x}^{(\mathrm{vis})*} \right)
\right].
\label{app:eq:sxy}
\end{equation}
In order to suppress the statistical error, we sum the values of
Eq.~(\ref{app:eq:sxy}), sampled at some interval in a single run, and
divide the sum by the number of samples after the run is terminated.
We have verified that the errors between independent runs are negligible.

\bibliography{./Suzuki-Hayakawa}

\end{document}